\documentstyle[tighten,preprint,aps]{revtex}

\def\e{\epsilon}

\def\lll{\lambda}

\def\R{R^{(2)}}

\newcommand{\be}{\begin{equation}}
\newcommand{\ee}{\end{equation}}

\input epsf

\begin{document}
\preprint{WISC-MILW-96-TH-13}
\title{Predictability and semiclassical approximation 
at the onset of black hole formation}
\author{Sukanta Bose,\footnote{Electronic address: {\em bose@csd.uwm.edu}}  
Leonard Parker,\footnote{Electronic address: {\em leonard@cosmos.phys.uwm.edu}} 
and Yoav Peleg\footnote{Electronic address: {\em yoav@csd.uwm.edu}} 
}
\address{Department of Physics \\
University of Wisconsin-Milwaukee, P.O.Box 413 \\
Milwaukee, Wisconsin 53201, USA}
\maketitle
\vskip 0.5cm
\begin{abstract}
We combine analytical and numerical techniques to study the collapse 
of conformally coupled massless scalar fields in semiclassical 
2D dilaton gravity, with emphasis on solutions just below criticality 
when a black hole almost forms. We study classical information and 
quantum correlations. We show explicitly how recovery of information 
encoded in the classical initial data from the outgoing 
classical radiation becomes more difficult as criticality is approached. 
The outgoing quantum radiation consists of a 
positive-energy flux, which is essentially 
the standard Hawking radiation, followed by 
a negative-energy flux which ensures energy conservation 
and guarantees unitary evolution through 
strong correlations with the positive-energy Hawking radiation. 
As one reaches the critical solution there is a breakdown of unitarity. 
We show that this breakdown of predictability is intimately related to  
a breakdown of the semiclassical approximation.
\end{abstract}
\vspace{1.0cm}

\newpage

\section{Introduction}
Quantum radiation from black holes \cite{Hawking} is necessary in order to 
maintain the consistency of the second law of thermodynamics with the 
existence of black holes \cite{Beken}. On the other hand, evaporation 
of the black hole reveals one of the most fundamental problems 
in theoretical physics; the question of unitary evolution of 
the Universe. Does evolution from an initial pure state take place 
non-unitarily to a final mixed state \cite{Haw2}, or unitarily to 
a final pure state \cite{tHooft} ? 
One of the major obstacles to a better understanding 
of the Hawking effect is the complexity of four-dimensional (4D) 
semi-classical gravity \cite{Birrell}. Simplified models which 
may give insight into the possible answers are two-dimensional (2D) 
dilaton quantum gravity theories \cite{Harvey}. 
The dilaton field, viewed as part 
of the geometrical structure restores dynamics in 2D analogous to that 
of spherically symmetric 4D Einstein gravity. We consider the 
formation and evaporation of a 2D 
black hole by the collapse of massless matter scalar fields. 
The evaporation of the black hole via production of quanta of the 
matter fields can be fully traced in the 2D semiclassical theory, 
including the back-reaction of the evaporation on the geometry 
\cite{deAlwis,Bilal,RST,BPP1,Cruz}. 

If the energy and energy density of the infalling matter are 
sufficiently large, then the incoming matter forms a black hole. 
Otherwise, the original incoming matter escapes to infinity 
and no black hole is formed. In this latter case the evolution is unitary 
and no information is lost. These unitary solutions of the semiclassical 
theory are called subcritical solutions 
\cite{ChungVerlinde,Thorlacius,DasMukherji,BPP2}. 
The study of subcritical solutions just below the 
critical threshold in which a black hole is formed may help us 
to understand the process of semiclassical black hole formation 
and its influence on information. Moreover, the interesting results 
obtained in classical gravity concerning critical behavior 
at the onset of classical black hole formation \cite{Choptuik} 
make it important to examine this critical behavior in the
context of semiclassical physics. In the critical regime of the 
4D Schwarzschild black hole, the black hole mass approaches zero 
and the curvature near the horizon becomes large, so semiclassical 
effects must be considered. Since the 4D semiclassical theory is 
quite complicated, insight may be gained by considering a 2D theory 
that shares many of its dynamical features, namely, 2D dilaton 
gravity \cite{Harvey}. The 2D dilaton models 
can be derived from 4D almost extremal 
dilatonic black holes \cite{Gibbons} 
using the Kaluza-Klein reduction \cite{Peleg}. 

Although the general properties of the 2D subcritical solutions, 
namely that they are stable and unitary, have been known for sometime
\cite{ChungVerlinde,Thorlacius,DasMukherji}, 
investigation of the explicit evolution reveals some new features 
of physical importance \cite{BPP2}. 
In Ref. \cite{BPP2} we studied the evolution of the subcritical 
solutions with infalling matter in the form of shock-waves. In this work 
we extend our study to include the behavior of subcritical 
solutions for general smooth initial data, with emphasis on the  
near-critical solutions. 

In Sec. II we present our model of 2D semiclassical dilaton gravity as 
an initial value problem. We also derive the general equations to be 
integrated numerically for arbitrary initial data. 
In Sec. III we give examples involving smooth infalling matter. 
The previous shock-wave results \cite{BPP2} appear as a limiting case of 
these examples.

In Sec. IV we address the question of information.
The information that may be lost in the process of black hole evaporation 
is related to the correlations between the outside world and the 
interior of the black hole. Two types of information are involved: 
(i)``Classical information'', carried by the classical matter that 
forms the black hole, and  
(ii) ``Quantum information'', encoded in  
quantum correlations between outgoing and incoming pairs of particles 
created by the collapse geometry. 
A quantity that plays an important role in 
understanding the structure of the subcritical solutions is the outgoing 
radiation reaching future asymptotic null infinity, $\Im^+_R$. In previous 
work \cite{BPP2} an explicit form of that radiation was found for the 
first time in semiclassical gravity. This outgoing radiation is intimately 
related to the question of information. In principle, for the subcritical 
solutions one should be able 
to recover the complete information given by the initial data from that 
outgoing radiation. We show how in the subcritical solutions, 
classical and quantum information is encoded in the 
outgoing radiation reaching asymptotic future null infinity. 

In Sec. V we consider the subcritical solutions just below criticality.
We show that as in the classical case, also in the semiclassical case 
solution space can be divided continuously into two regions, i.e., there 
exist continuously varying parameters, $p_i$, 
in solution space, such that for $p_i < p^*_i$ the 
evolved scalar field will not form a black hole (the subcritical solutions), 
while for $p_i > p^*_i$ a black hole will be formed (the supercritical 
solutions). We show that as the critical solution is approached ($p_i 
\rightarrow p_i^*$) the outgoing energy flux diverges and the fluctuations 
in the outgoing energy density become very large, implying a 
break down of the semiclassical approximation at criticality. 

In Sec. VI we show that near criticality the density of information 
encoded in the outgoing radiation reaching $\Im^+_R$ becomes very large
and diverges at criticality. This divergence results in an 
apparent  breakdown of predictability that coincide with the  
breakdown of the semiclassical approximation. We present our conclusions 
in Sec. VII. 

\section{The model}
\subsection{1-loop effective action}

Recently we have proposed a modified theory of 2D semiclassical dilaton 
gravity \cite{BPP1}. The effective action of the modified theory is 
  \be \label{Seff}
S_{\mbox{eff}} = S_{CGHS} + N S_{PL} + S_{\mbox{corr}} , 
  \ee
where $S_{CGHS}$ is the Callan-Giddings-Harvey-Strominger (CGHS) 
classical action 
\cite{CGHS}, 
  \be \label{CGHS}
S_{CGHS} = {1\over 2\pi} \int d^{2}x \sqrt{-g}
\left[ e^{-2\phi} \left( \R + 4 (\nabla \phi)^{2} + 4 \lambda^2 \right)
- {1\over 2} \sum_{i=1}^{N} (\nabla f_{i})^{2} \right] , 
  \ee
$S_{PL}$ is the Polyakov-Liouville action \cite{Polyakov} 
that incorporates the 1-loop corrections corresponding to the trace 
anomaly of the stress energy momentum tensor of each of the $N$ 
quantum matter fields,  
  \be \label{PL}
S_{PL} = - {\hbar \over 96\pi}
\int d^{2}x \sqrt{-g(x)}
\int d^{2}x' \sqrt{-g(x')} \R(x) G(x,x') \R(x') , 
  \ee
and 
  \be \label{Scorr}
S_{\mbox{corr}} = {N\hbar\over 24\pi} \int d^{2}x \sqrt{-g} \left(
(\nabla \phi)^{2} - \phi \R \right) 
  \ee
is a local counter-term that we add in order to get an exactly solvable 
theory. In the above $\phi$ is the dilaton field, $\R$ is the 2D Ricci scalar,
$\lambda$ is a positive constant, $\nabla$ is the covariant derivative,
and $G(x,x')$ is an appropriate Green's function for $\nabla^2$. 
The $N$ real value functions, $f_i(x)$, are the classical values of the 
massless scalar fields. One can regard each of the $f_i(x)$ as the 
expectation value of the quantum field operator, $\hat{f}_i(x)$, in an 
appropriate quasi classical coherent state, $|\alpha \rangle$ \cite{Preskill}. 
The effective action (\ref{Seff}) describes the full quantum theory 
in the large $N$ limit, in which case the fluctuations of $\phi$ and 
$g_{\mu \nu}$ can be neglected \cite{ChungVerlinde,BPP1}. 
Recently Mikovi\'{c} showed that one can derive the effective action 
of Eq. (\ref{Seff}) from $S_{CGHS}$ by fixing the diffeomorphism gauge,
solving the constraints, and then 
quantizing the reduced system. After choosing an appropriate initial 
quantum state, one recovers the action (\ref{Seff}) as a 1-loop effective 
action \cite{Mikovic}. 

In null coordinates, $z^{\pm}$, and conformal gauge, $g_{++}=g_{--}=0$, 
$g_{+-}=-(1/2)\exp(2\rho)$, the action (\ref{Seff}) takes the form 
  \begin{eqnarray} \label{Scon}
S_{\mbox{eff}} = {1\over \pi} \int dz^+ dz^- &\bigglb[& (\partial_-Y) 
\partial_+ (X - {\kappa\over 2} Y) + (\partial_+Y) \partial_- 
(X - {\kappa\over 2} Y) + \lambda^2 \exp(-2Y) \nonumber \\
&+& {1\over 2} \sum_{i=1}^{N} \partial_+ f_i 
\partial_+ f_i \biggrb] ,
  \end{eqnarray}
where $X \equiv \exp(-2\phi)$, $Y \equiv \phi - \rho$, and $\kappa = 
N \hbar / 12$. In the large $N$ limit we take $\hbar$ to approach zero 
while keeping $\kappa$ finite. 
The kinetic action density of the system described in Eq. 
(\ref{Scon}) is a bilinear symmetric form $(\partial_+ \Theta ) M(\phi) 
(\partial_- \Theta )$, where $\Theta$ is a vector comprised of the 
$(N + 2)$ fields $X$, $Y$ and the $N$ matter fields, $f_i$, and $M(\phi)$ 
is an $(N+2)\times (N+2)$ symmetric matrix. One can verify that the 
determinant of $M$ is proportional to $X^2 = \exp(-4 \phi)$, and unlike in 
other models of modified dilaton gravity \cite{deAlwis}, 
here this determinant is 
non-vanishing for all real values of $\phi$. The vanishing of the 
determinant at $X(x^+,x^-) = 0$ signals a singularity.
 
The equations of motion 
derived from varying the action (\ref{Scorr}) with respect to 
$X$, $Y$ and $f_i$ are 
  \begin{eqnarray} 
\partial_+ \partial_- X &=& -\lambda^2 \exp(-2Y) , \label{Xeq} \\
\partial_+ \partial_- Y &=& 0 , \label{Yeq} \\
\partial_+ \partial_- f_i &=& 0 . \label{feq} 
  \end{eqnarray} 
The constraints (from varying the action (\ref{Seff}) with respect to 
$g_{\pm \pm}$) are 
  \be \label{constraints}
-\partial^2_{\pm} X - 2 \partial_{\pm} X \partial_{\pm} Y 
- T^{cl}_{\pm \pm} + \kappa \left[ (\partial_{\pm} Y )^2 
+ \partial^2_{\pm} Y + t_{\pm}(z^{\pm}) \right] = 0 , 
  \ee
where after varying with respect to $g_{\mu \nu}$ ($\mu=\pm$, $\nu=\pm$) 
we set $g_{++}=g_{--}=0$ to get (\ref{constraints}).   
Here $T^{cl}_{\pm \pm} = (1/2) \sum_{i} (\partial_{\pm} f_i )^2 $ 
is the classical contribution to the energy-momentum tensor of the 
$N$ matter fields, and $t_{\pm}(z^{\pm})$ are integration functions 
determined by the specific quantum state of the matter scalar fields. 
A conformal coordinate transformation of the form $z^+ \rightarrow y^+$ and 
$z^- \rightarrow y^-$, preserves the form of the metric, $g_{\pm \pm}=0$ 
and $g_{+-}=-(1/2)\exp(2 \bar{\rho})$, 
where the new conformal mode function is 
related to the old one by $\bar{\rho}(y^+,y^-)=\rho(z^+,z^-) + 
\ln\sqrt{dz^+/dy^+} + \ln\sqrt{dz^-/dy^-} $. 
The dilaton field is a scalar, and therefore the field $Y$ transforms like 
$- \rho$. The general solution of Eq. (\ref{Yeq}) for $Y$ is   
$Y(z^+,z^-) = Y_+(z^+) + Y_-(z^-)$, and we can choose the 
coordinates $x^{\pm} = \int^{z^{\pm}} \exp[-2Y_{\pm}(\tilde{z}^{\pm})] 
d\tilde{z}^{\pm}$,  
for which $Y(x^+,x^-) \equiv 0$. In the following we use 
these ``Kruskal'' coordinates, denoted by $(x^+,x^-)$, for which 
$\phi(x^+,x^-)=\rho(x^+,x^-)$.  Henceforth, the indices $\pm$ on tensors
will refer to components in the Kruskal coordinates.

In the Kruskal gauge the general solutions to (\ref{Xeq}) and (\ref{feq}),  
subject to the constraints (\ref{constraints}), are 
  \begin{eqnarray} 
X(x^+,x^-) &=& -\lambda^2 x^+ x^- - \int^{x^+} dx^+_2 \int^{x^+_2} dx^+_1 
\left[ T^{cl}_{++}(x^+_1) - \kappa t_+(x^+_1) \right] \nonumber \\
& & - \int^{x^-} dx^-_2 \int^{x^-_2} dx^-_1 
\left[ T^{cl}_{--}(x^-_1) - \kappa t_-(x^-_1) \right] 
\label{Xsolution} 
  \end{eqnarray}
and 
  \be
f_i(x^+,x^-) = f_i^+(x^-) + f_i^-(x^-) . \label{fsolution}
  \ee

\subsection{The initial value problem}
\begin{figure}
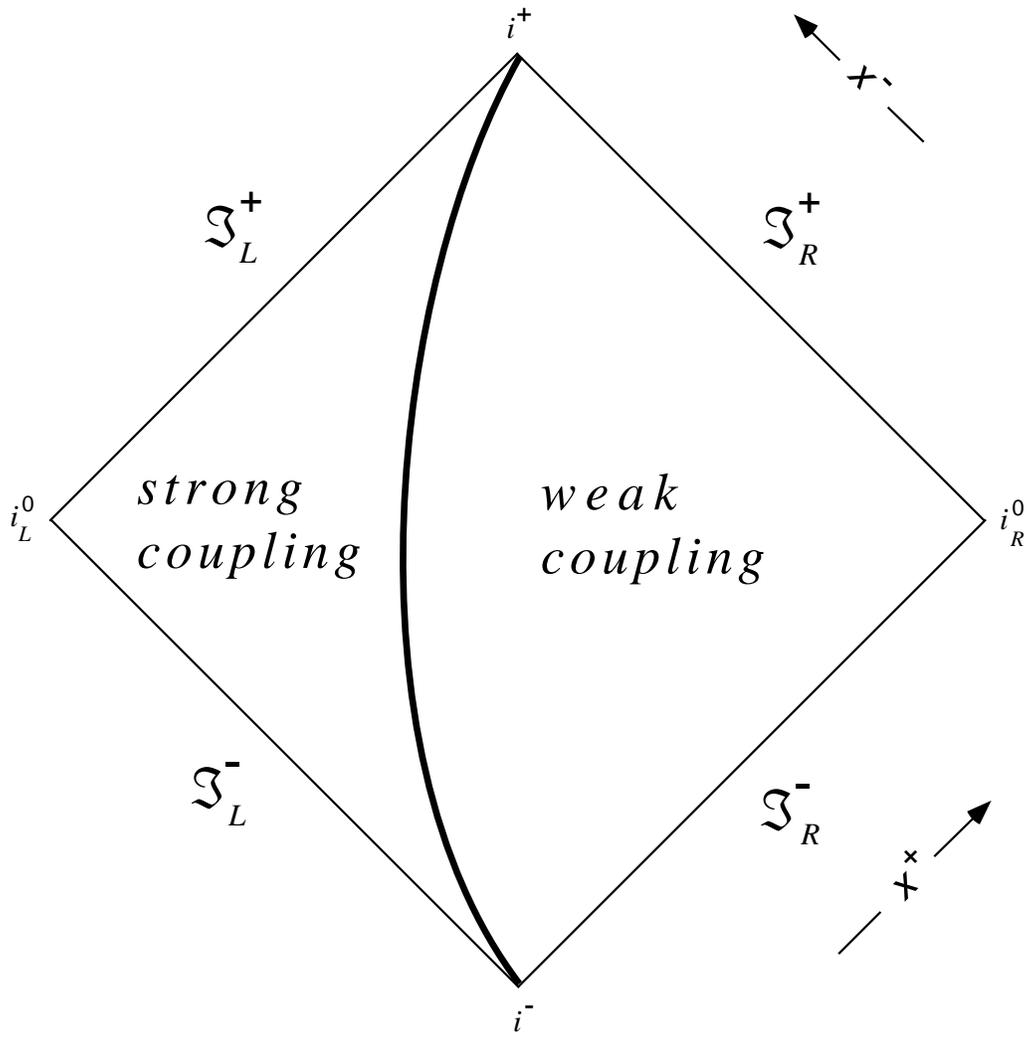
 
\caption{Penrose diagram of the Linear Dilaton solution. The heavy curve 
is the boundary curve separating the regions of weak and strong coupling.} 
\label{fig1}
  \end{figure}

Consider first the linear dilaton (LD) solution which 
corresponds to $f_i(x^+,x^-) = 0$ and $t_{\pm}(x^{\pm})=0$ in 
(\ref{Xsolution}), namely, the 
solution $X_{LD}(x^+,x^-) = -\lll^2 x^+ x^-$. It is defined for 
$0 < x^+ < \infty$ and $-\infty < x^- < 0$. In the 
manifestly flat coordinates, $\sigma^{\pm} \equiv \tau \pm \sigma = 
\pm \lll^{-1} \ln(\pm \lll x^{\pm})$, the LD solution corresponds 
to the flat metric, $ds^2 = -d\sigma^+ d\sigma^-$, and the dilaton field 
has the linear form, $\phi = -\lll \sigma$. 
As shown in Fig.\ \ref{fig1}, the null curve $x^+=0$ defines left asymptotic 
past null infinity, $\Im^-_L$, the null curve $x^-= -\infty$ defines right 
asymptotic past infinity, $\Im^-_R$. The null curve $x^-=0$ is left 
asymptotic future infinity, $\Im^+_L$ and $x^+=+\infty$ is right 
asymptotic future infinity, $\Im^+_R$. 
In general, the initial data on $\Im^-_{L}$ and $\Im^-_R$ 
determine completely the solution in the region $0 < x^+ < \infty$, 
$-\infty < x^- < 0$. Specifying these initial data is equivalent to giving 
$f_i^-(x^-)$, and $t_-(x^-)$ on $\Im^-_L$ and 
$f_i^+(x^+)$ and $t_+(x^+)$ on $\Im^-_R$. Giving these functions, 
we integrate (\ref{Xsolution}) to find the solution everywhere, 
however such a solution may not be physically acceptable in the whole 
space-time, since singularities may appear. 
One can say that our 1-loop effective theory is 
exactly solvable however there is one major difficulty with such 
an approach. Consider first the LD solution: 
while on $\Im^-_R$ we have $\sigma \rightarrow \infty$  
and $\exp(2\phi) = 0$, on $\Im^-_L$ ($\sigma \rightarrow -\infty$) 
we have $\exp(2\phi) \rightarrow \infty$. From the action (\ref{CGHS}) 
we see that 
$\exp(2\phi)$ plays the role of the ``coupling constant'', and 
so the coupling {\em diverges} on $\Im^{\pm}_L$. One can split the 
space-time into a region of weak coupling and a region of strong 
coupling, see Fig.\ \ref{fig1}. Those two regions are divided by a curve, 
the ``boundary curve'' (specified below in more detail). 
%
In the strong coupling region we cannot 
trust the 1-loop effective theory, especially on $\Im^{\pm}_L$. 
Therefore giving the initial data on $\Im^-_L$ is in general 
unphysical. One way to avoid this problem is to 
consider the solutions only in the weak coupling region and impose on a 
time-like curve boundary conditions that preserve unitarity and conserve 
energy. These criteria are satisfied by imposing  
reflecting boundary conditions on the 
boundary curve. Thus, the initial value problem that we 
define is the following: on $\Im^-_R$ we give the initial data, 
$f_i^+(x^+)$ and $t_+(x^+)$, and on the boundary curve  
defined by $x^+=x^+_B(x^-) \equiv p(x^-)$, 
we impose the reflecting boundary condition.

Before specifying the boundary condition we elaborate on the 
energy-momentum tensor. The c-number $T^f_{\mu \nu}$, which one gets 
after varying the effective action in (\ref{Seff}) with respect to the 
metric $g^{\mu \nu}$, can be regarded as the expectation value of 
the quantum operator, $\hat{T}^f_{\mu \nu}$, in the quasi-classical 
coherent state, $|\alpha \rangle$, \cite{Preskill}. Namely, 
  \begin{eqnarray*}
T^f_{\mu \nu} \equiv 
\langle \alpha | \hat{T}^f_{\mu \nu} | \alpha \rangle = T^{cl}_{\mu \nu} 
+ \langle T_{\mu \nu} \rangle \; , 
  \end{eqnarray*}
where in the expansion of $\langle \alpha | \hat{T}^f_{\mu \nu} 
| \alpha \rangle $ the first term, $T^{cl}_{\mu \nu}$, is of order 
$\hbar^0$ and is the classical part of the stress tensor, while 
the second term, $\langle T_{\mu \nu} \rangle$, is 
of order $\hbar$ (or $N \hbar$ for $N$ fileds) and is the one-loop contribution. 
The one-loop contribution to the energy-momentum 
tensor is given by 
  \be \label{tquant}
\langle T_{z^{\pm} z^{\pm}} \rangle = \kappa [ \partial^2_{z^{\pm}} \rho 
- (\partial_{z^{\pm}} \rho)^2 - t_{z^{\pm}}(z^{\pm}) ] \qquad ; \qquad 
\langle T_{z^+ z^-} \rangle = \kappa \partial_{z^+} \partial_{z^-} \rho \; , 
  \ee 
in some general null coordinates
$(z^+, z^-)$. Equation (\ref{tquant}) follows by integrating 
$\nabla_\mu \langle T^{\mu \nu} \rangle = 0$ and using the trace anomaly 
\cite{Deser} $\langle T^{\mu}_{\mu} \rangle = - \kappa R^{(2)} / 2$ for 
the $N$ massless matter fields $\hat{f}_i$ \cite{Davies}, 
or equivalently by varying the Polyakov-Liouville 
action (\ref{PL}) with respect to the metric \cite{Polyakov}.  

One can argue that operationally 
the split between the classical and quantum radiation is not well-defined when 
we consider a single (scalar) field. However, for $N$ fields 
one could classically excite only some of the fields. Then the total 
radiation in the remaining fields is just the quantum part. To 
operationally distinguish 
among the $N$ fields, one can, for example, add another 
quantum number to these scalar fields.  

The reflecting boundary condition is \cite{Fulling,Thorlacius,Birrell,BPP2}
  \be \label{reflect}
T^f_{--}(x^-) = (p'(x^-))^2 T^f_{++}(x^+_B(x^-)) + \kappa 
(p'(x^-))^{1/2} \partial^2_- (p'(x^-))^{-1/2}
  \ee
where $ ' = \partial / \partial x^-$. The last expression on the 
right hand side (r.h.s.) of 
Eq. (\ref{reflect}) is due to quantum particle creation from 
the boundary, which is effectively a moving mirror \cite{Birrell,Fulling}. 
One can split the reflecting boundary condition (\ref{reflect}) 
into its classical and 1-loop parts, 
  \begin{eqnarray}
T^{cl}_{--}(x^-) &=& 
(p'(x^-))^2 T^{cl}_{++}(x^+_B(x^-)) \label{classb} \\ 
t_-(x^-) &=& (p'(x^-))^2 t_+(x^+_B(x^-)) + 
{(p'(x^-))^{1/2} \partial^2_- (p'(x^-))^{-1/2}\over 
1 + \kappa/ (2X_B) } \label{quantb}
  \end{eqnarray}
If we take the fields $f_i$ to satisfy Neumann or Dirichlet boundary 
conditions, then the classical reflecting boundary condition 
(\ref{classb}) is satisfied.


Next we consider the boundary curve. We would like it to separate 
the regions of weak coupling , $\exp(2\phi) < g^2_c$, and strong coupling, 
$\exp(2\phi) > g^2_c$, where $g^2_c$ is some parameter that 
specifies the value of the coupling below which the 1-loop effective 
theory is trustworthy. Since $X(x^+,x^-) = \exp(-2\phi)$, we can 
define the boundary curve to be the curve on which 
$X(x^+,x^-) = X_B \equiv g^{-2}_c = \mbox{const}$.  
The solution $X(x^+,x^-)$ should be determined by the initial data on 
$\Im^-_R$ and by the boundary 
condition (\ref{reflect}). The boundary curve, $X(x^+,x^-)=X_B$, 
depends on the solution $X(x^+,x^-)$. Thus, we get a highly non-linear 
problem, unlike the straightforward problem, with initial data given 
on $\Im^-$, having solutions (\ref{Xsolution}) and (\ref{fsolution}). 
We next reduce this non-linear problem to solving a single ordinary 
differential equation.

\subsection{The boundary equation as a second order ODE}
The function $T^{cl}_{++}(x^+)$ can be viewed as the part of the 
initial data on $\Im^-_R$ that describes 
the classical profile of the infalling matter. We take 
$T^{cl}_{++}(x^+)$ to be a general function of $x^+$ with  
compact support. On the other hand, $t_+(x^+)$ describes the quantum state 
on $\Im^-_R$. Since in this work we would like to 
study the Hawking effect, we take the quantum state to be such that 
on $\Im^-_R$ we have no quantum radiation, (i.e.,
$\langle T_{vv} \rangle (v) = 0$, 
where $v$ is the asymptotically 
flat null coordinate on $\Im^-_R$, $v = \lll^{-1} \ln(\lll x^+)$). 
Since in the asymptotically flat null coordinates the conformal mode, $\rho$, 
approaches zero on $\Im^-_R$, we get from (\ref{tquant}) that in these 
coordinates $t_v (v) = 0$. To find the corresponding $t_+ (x^+)$ in 
Kruskal coordinates, we use the tensor transformation of 
$\langle T_{z^{\pm} z^{\pm}} \rangle$  in Eq. (\ref{tquant}) 
(under a conformal coordinate transformation) and get
\be \label{trelations}
t_+ (x^+) = {\left( \partial v \over \partial x^+
\right)}^{2} \left( t_v (v) - {1\over 2} 
D^{S}_{v} [x^+] \right) = {1\over (2x^+)^{2}}
\ \ ,
\ee
where $D^{S}_{y}[z]$ is the Schwarz operator defined as
\be \label{Schwarz}
D^{S}_{y}[z] = 
(\partial^{3}_{y} z) / (\partial_{y} z) - {3\over 2} 
{\left( \partial^{2}_{y} z / \partial_{y} z \right)}^{2}
\ee
and we use $t_{v}(v)=0$. Eq. (\ref{quantb}) then becomes 
  \be \label{tcondition1}
t_-(x^-) = {1\over 4} {\left( {p'(x^-) \over p(x^-)} 
\right)}^2 + {(p'(x^-))^{1/2} \partial^2_- (p'(x^-))^{-1/2}
\over 1 + \kappa/ (2X_B)} \; . 
  \ee
The second term on the r.h.s. of Eq. (\ref{tcondition1}) is the result 
of particle creation from the dynamical boundary. 
Recall that in the 4D Hawking 
effect \cite{Hawking}, the ``boundary curve" is $r=0$ (the fixed point 
of spherical symmetry). The curve $r=0$ does not act as a moving mirror
and there is no particle creation at $r=0$. The creation of particles 
(i.e., the Hawking radiation) is due to the curvature of space-time near 
the horizon. Can we also eliminate the moving 
mirror effect in our 2D theory? We see from (\ref{tcondition1}) that if 
$X_B << \kappa$, then the moving mirror term is negligible.  
This is consistent with the fact that $X = \exp(-2\phi)$ is 
indeed the 4D radial coordinate \cite{Peleg,RST1}, and $X_B \rightarrow 0$ 
corresponds to $r \rightarrow 0$ in 4D. However, one should be careful when  
taking the limit $X_B \rightarrow 0$. It corresponds to the limit 
$g^2_c \rightarrow 
\infty$, which defines the strong coupling region. The 1-loop effective 
theory (\ref{Seff}) is trustworthy as long as $g^2_c \hbar << \kappa << 1$. 
In this case (in the region of interest, i.e., $X(x^+,x^-) \geq X_B$) 
the quantum corrections for the dilaton-gravity part are 
negligible compare to the 1-loop corrections for the $N$ 
scalar fields, which in turns are small compared to the 
classical contribution. 
Therefore, we need  to satisfy both $g^2_c \hbar << 
\kappa << 1$ and $X_B << \kappa$. With $X_B^{-1} = g_c^2$, we combine the 
conditions and get  
  \be \label{largeNlimit}
{1\over N} << X_B << N \hbar << 1 . 
  \ee
One can always take the large $N$ limit in such a way that (\ref{largeNlimit}) 
is satisfied and $X_B$ is arbitrarily close to zero. 
In the following we take this large $N$ limit. Eq. (\ref{tcondition1}) 
then reduces to 
  \be \label{tcondition}
t_-(x^-) = {1\over 4} {\left( {p'(x^-) \over p(x^-)} 
\right)}^2 . 
  \ee

One can write the general solution (\ref{Xsolution}) in the form 
  \be \label{Xsolution2}
X(x^+,x^-) = -\lll x^+ (\lll x^- + \lll^{-1} P_+(x^+) ) - {\kappa \over 4} 
\ln(\lll x^+) + {M(x^+) \over \lll} + F(x^-) ,
  \ee
where $M(x^+)$ and $P_+(x^+)$ are the mass and momentum of the 
infalling classical matter, 
  \be \label{MandP}
M(x^+) \equiv \lll \int_0^{x^+} x^+_1 T^{cl}_{++}(x^+_1) dx^+_1 
\qquad , \qquad P_+(x^+) \equiv 
\int_0^{x^+} T^{cl}_{++}(x^+_1) dx^+_1 
\ \ ,
\ee
and
\be \label{F}
F(x^-) \equiv  - \int^{x^-} dx^-_2 \int^{x^-_2} dx^-_1 
\left[  T_{--}^{cl}(x^-_1) - \kappa t_-(x^-_1) \right] \; . 
  \ee
Here $F(x^-)$ is a function of $x^-$ 
to be determined by the boundary conditions. Before we obtain the
equation for the boundary curve, let us define the following dimensionless
quantities:
\be \label{dimlesscoord}
z \equiv \lll x^- \qquad
\mbox{and} \qquad q(z) \equiv \lll x^+_B(x^-(z)) = \lll p(x^-(z))
\ \ ,
\ee 
where $q(z)$ is a dimensionless function that specifies the location of
the boundary curve.
Using (\ref{constraints}), (\ref{tcondition}), (\ref{Xsolution2})   
and acting with 
$\partial^2 / \partial z^2$ on the boundary equation $X(x^+_B,x^-) = X_B$, 
we get a second order ODE for $q(z)$  
  \be \label{ODE}
\left[ z + P_q(q) + {\kappa \over 4 q(z)} \right] {d^2 q\over dz^2} 
+ 2 {dq\over dz} 
+ 2 \left[ T^{cl}_{qq}(q) - {\kappa\over 4 q^2(z)} \right] 
{\left( {dq\over dz} \right)}^2 = 0 , 
  \ee
where 
\be \label{Tqq}
T^{cl}_{qq}(q) \equiv \lll^2 T_{++}(x^+_B(x^-)) \qquad 
\mbox{and} \qquad P_q(q)=\int^q T^{cl}_{qq}(\bar{q})d\bar{q}
\ee 
are given functions of $q$, which are determined by the initial data 
on $\Im^+_R$. Eq. (\ref{ODE}) is a second order, non-linear, ordinary 
differential equation for the boundary curve $q(z)$. 
We solve it numerically for different profiles of the infalling matter
using an embedded fifth-order Runge-Kutta 
ODE integration routine \cite{Press}. After obtaining $q(z)$ we use 
Eqs. (\ref{classb}) and (\ref{tcondition}) to find $T_{--}^{cl}(x^-)$ 
and $t_-(x^-)$ and integrate Eq. (\ref{Xsolution}) to obtain the 
solution $X(x^+,x^-)$.

\section{Solutions}
\subsection{Vacuum solutions}
In this subsection we consider solutions of (\ref{ODE}) with different 
profiles of infalling null matter. First consider the vacuum solutions, 
for which $T^{cl}_{qq}(q) = P_q(q) = 0$. 
The general solution of (\ref{ODE}) with 
$T^{cl}_{qq}(q) = P_q(q) = 0$ and the initial conditions 
$q(z \rightarrow -\infty) \rightarrow 0$ and  
$(dq/dz)(z \rightarrow -\infty) \rightarrow 0$, is 
  \be \label{vacuumboundary}
q_{vac}(z) = - {a\over z} ,
  \ee
where $a$ is some positive constant. One finds from (\ref{MandP}) and 
(\ref{vacuumboundary}) the one-parameter family of vacuum solutions 
  \be \label{vacuumsolution}
X_{vac}(x^+,x^-) = -\lll^2 x^+ x^- - {\kappa\over 4} \ln(-\lll x^+ x^-) 
+ {\kappa \over 4} \ln(a) - a . 
  \ee
These are static solutions \cite{BPP1}. For $a > \kappa/4$ the solution 
(\ref{vacuumsolution}) has a time-like singularity in the strong coupling 
region, $e^{2\phi} \rightarrow \infty$, while for $a < \kappa/4$ the 
solution has null singularities in the weak coupling region $e^{2\phi} 
\rightarrow 0$. The solution with $a = \kappa/4$ 
is everywhere regular, has the geometry of a semi-infinite 
throat, and can be regarded as the ground state of the theory \cite{BPP1}. 
One can show that ~\cite{Bilal-ADM} compared to this ground state, the ADM 
mass of the static solution in Eq. (\ref{vacuumsolution}) is 
\be \label{vacADM}
M = \lll \left[ {\kappa \over 4} \ln \left( {4a\over \kappa} \right) - 
                                     \left( a -{\kappa \over 4}\right)
         \right]
\ \ .
\ee
 
\subsection{Smooth infalling matter}

\begin{figure}
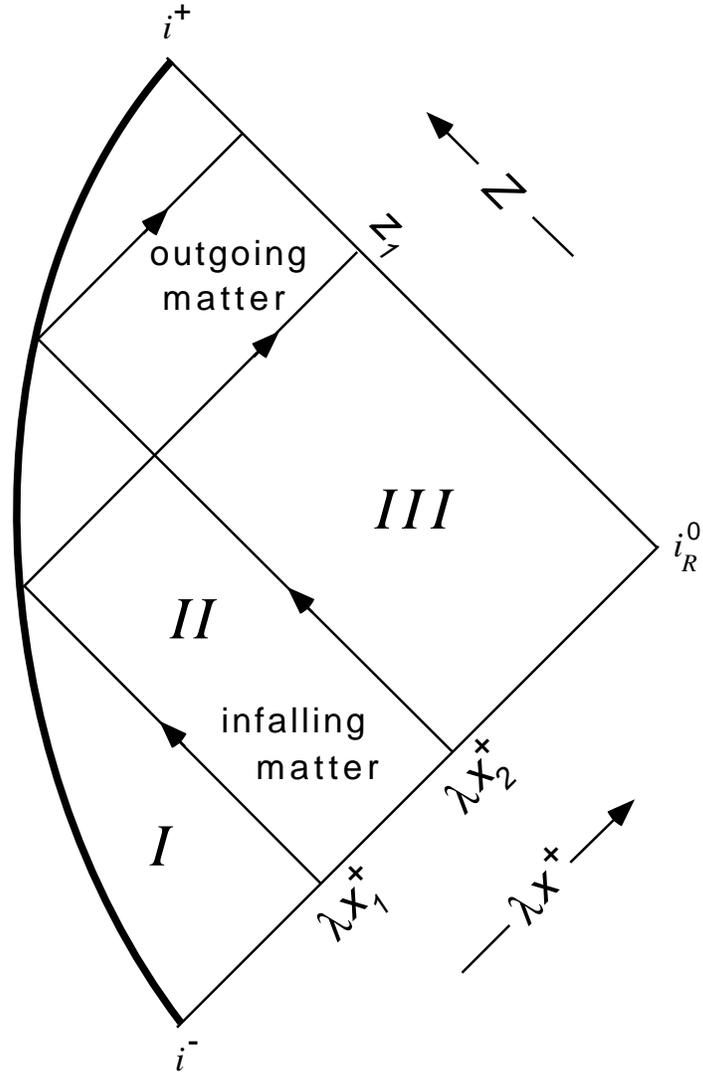
 
\caption{Penrose diagram of a typical subcritical solution. The infalling 
matter with support on the interval $x^+_1 < x^+ < x^+_2$ on $\Im^-_R$, 
is reflected from the boundary to become outgoing toward future 
null infinity, $\Im^+_R$. Here $z = \lll x^-$.} 
\label{fig2}
  \end{figure}

In the following we consider more general infalling matter with compact 
support, $x^+_1 < x^+ < x^+_2$. For $x^+ < x^+_1$, i.e., region I in 
Fig.\ \ref{fig2}, we have $T_{qq}^{cl} = P_q = 0$, 
and the solution is one of the vacuum solutions (\ref{vacuumsolution}). 
If the total mass of the infalling matter, $M \equiv M(x^+ \rightarrow 
\infty )$ in Eq. (\ref{MandP}), is above a critical value $M_{cr}$,
then a black hole is formed. The critical mass $M_{cr}$ is of the order of 
$\lll a $ \cite{BPP1} and is determined numerically from the profile of 
the infalling matter stress-tensor and the ADM mass of the initial spacetime.
We take $M/\lll \gg \kappa$ for the validity of the semiclassical 
approximation \cite{BPP3}.
Since we consider subcritical solutions in which a black hole
does not quite form, $a$ is at least of the order of $M/ \lll$ and hence
$a \gg \kappa$. We take $a=1$ and $\kappa = 10^{-6}$. Our results would be 
similar for any values of $a$ and $\kappa$ as long as $a \gg \kappa$. 
We also fix the scale of the $x^+$ coordinate such that $\lll x^+_1 = 1$. 
Define $z_1$ to be the value of $z = \lll x^-$ in Eq. (\ref{ODE})
corresponding to $q(z_1)=\lll x^+_1$ (see Fig.\ \ref{fig2}). 
Using $\lll x^+_1 = 1$, $a=1$ and 
$X_B \approx 0^+$ we find from Eq. (\ref{vacuumsolution}) that $z_1 = -1$.
For $z< z_1$, i.e, in regions I, II and III in Fig.\ \ref{fig2}, 
the solution of Eq. (\ref{ODE}) for $q(z)$ is given by 
Eq. (\ref{vacuumboundary}), and the solution for $X(x^+,x^-)$ is 
  \begin{eqnarray*} 
X(\lll x^- < z_1) = -\lll x^+[\lll x^- +\lll^{-1}P_+(x^+)] - {\kappa \over 4} 
\ln(-\lll^2 x^+ x^-) + {M(x^+)\over \lll} \; . 
  \end{eqnarray*}
For $z \geq z_1=-1$ (see Fig.\ \ref{fig2}) we have to integrate Eq. 
(\ref{ODE}) numerically to find the boundary curve $q(z)$.
So the range of $z$ to be integrated numerically is $[-1,0)$. 
The initial values for the numerical integration of the ODE (\ref{ODE}) are 
(using (\ref{vacuumboundary}) with $a=1$): 
$q(z=z_1) = 1$ and $dq/dz(z=z_1) = 1$.  

Before we give the results of our numerical integration, it is convenient
to define coordinates in which the dynamical metric of our spacetime 
corresponding to the solution (\ref{Xsolution2}) is manifestly asymptotically 
flat on $\Im_R^{\pm}$. In the Kruskal coordinates, the metric is
\be \label{metric2}
ds^2 = -\exp [2\rho (x^+, x^-)] dx^+ dx^- = - X^{-1} \> dx^+ dx^-
\ \ ,
\ee
where $X = \exp [-2\rho (x^+, x^-)]$ in these coordinates and is given by 
Eq. (\ref{Xsolution2}).
The required asymptotically flat coordinates are $(u, v)$, defined in terms
of the Kruskal coordinates by the following conformal coordinate 
transformation:
  \be \label{vu}
v=\lll^{-1} \ln(\lll x^+) \qquad \mbox{and} \qquad  
u = - \lll^{-1} \ln(-\lll x^- - P_+/\lll) \; , 
  \ee
where $P_+ = P_+(x^+ \rightarrow \infty)$ is the total momentum 
of the infalling matter. Rewriting the metric (\ref{metric2}) in terms
of the $(u, v)$ coordinates shows that $\exp[2\rho (v, u)] \to 1$
as $v \to \infty$ on $\Im_R^+$ and $u \to -\infty$ on $\Im_R^-$.

\begin{figure}
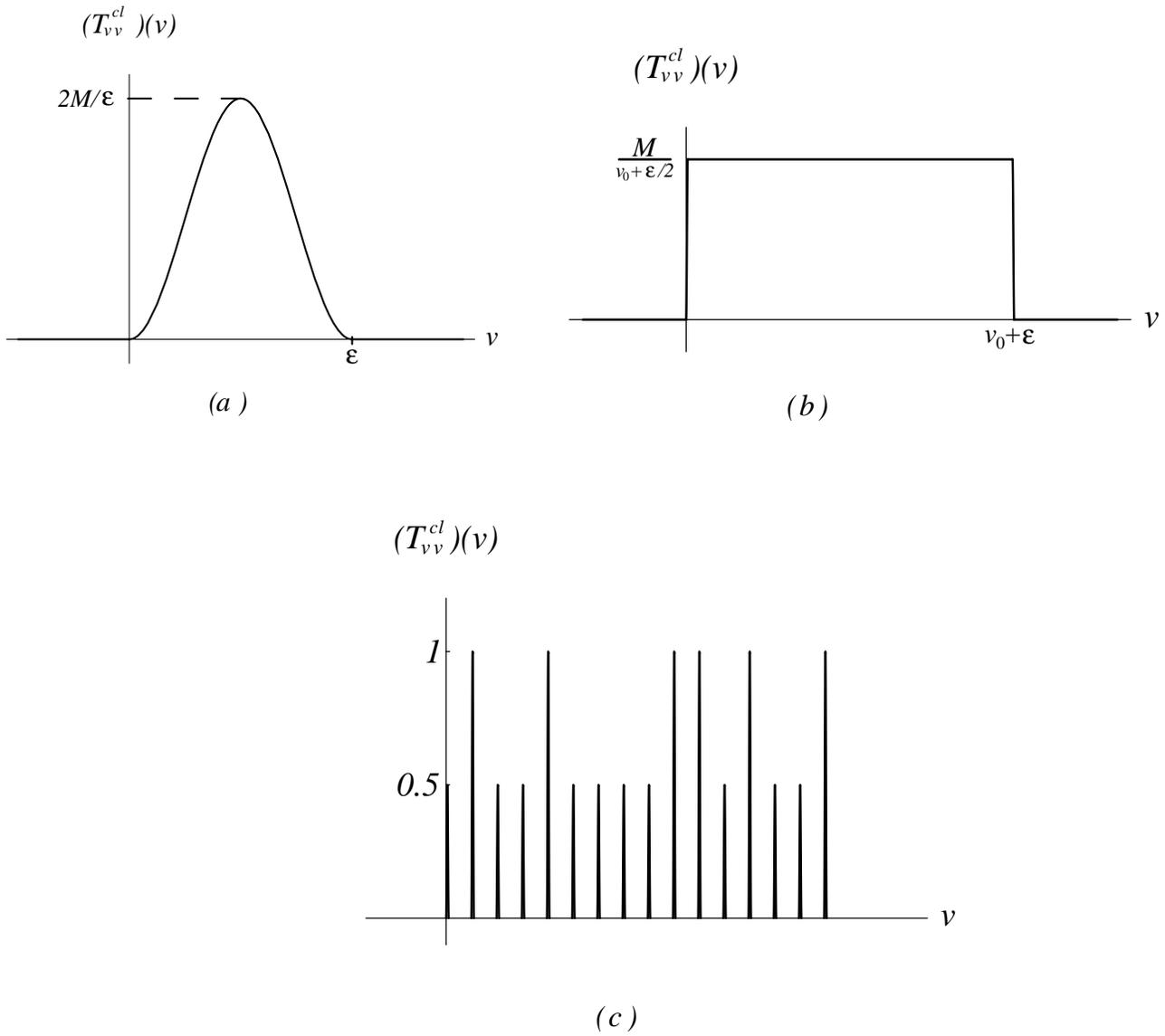
 
\caption{The incoming stress tensor, $T_{vv}(v)$, for three different 
families of initial data, with (a) corresponding to Eq. (32), 
(b) corresponding to Eq. (38) and (c) corresponding to Eq. (40).} 
\label{fig3}
  \end{figure}

Consider now the stress tensor describing a smooth profile of 
infalling matter:  
 \be \label{profile}
T^{cl}_{vv}(v) = \left\{ \begin{array}{cccc}
                          0 & , & & v < 0 \\
                      (2M/ \e) \sin^2(\pi v / \epsilon)
                            & , & & 0 < v <  \epsilon  \\
                          0 & , & & v > \epsilon \; , 
                          \end{array} \right. 
  \ee  
where $v = \lll^{-1} \ln (\lll x^+)$ . 
In Fig.\ \ref{fig3}a we show the profile (\ref{profile}). 
Here $M$ is the total mass of the infalling matter, 
$M = \int_0^{\infty} T^{cl}_{vv}(v) dv$, and $\epsilon$ is its width.   


To integrate (\ref{ODE}) we need to write $T_{qq}^{cl}$ in terms of $q$. 
Using (\ref{dimlesscoord}) and (\ref{Tqq}) in (\ref{profile}), we get 
  \be \label{qqprofile}
T^{cl}_{qq}[q] = \left\{ \begin{array}{cccc}
                          0 & , & & \ln(q) < 0 \\
          {2M \over \lll^2 \epsilon q^2} \sin^2(\pi \ln(q) / (\lll \epsilon))
                            & , & & 0 < \ln(q) < \lll \epsilon  \\
                          0 & , & & \ln(q) > \lll \epsilon \; . 
                          \end{array} \right. 
  \ee    
We numerically solve (\ref{ODE}) with (\ref{qqprofile}) 
for different values of $M/\lll$ and $\lll \e$, and 
find the corresponding boundary curves $q(z)$ and solutions $X(x^+,x^-)$. 

\begin{figure}
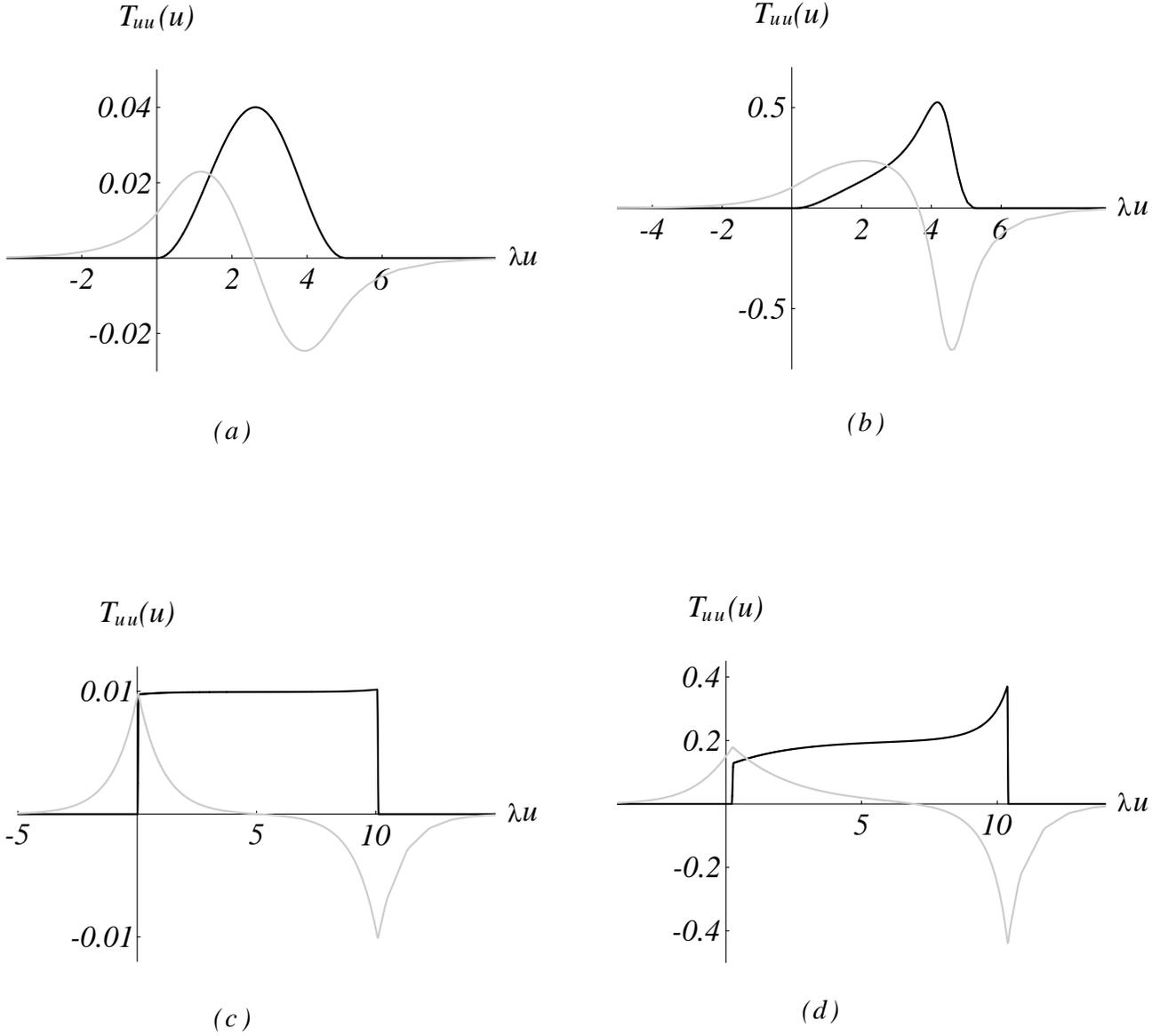
 
\caption{The stress tensor $T_{uu}$ in units of $\lambda^2$ describing the 
outgoing radiation on $\Im^+_R$. The solid curves 
describe the total outgoing radiation, $T^f_{uu}$, 
while the gray curves describe only the quantum part, $\langle T_{uu} \rangle$, 
scaled by $2/\kappa$. In (a) the mass is about 6\% 
and in (b) it is about 65\% of the critical mass for the infalling 
matter shown in Fig. 3(a). 
In (c) the mass is about 3\% and in (d) it is about 65\% of the critical mass 
for the case shown in Fig. 3(b).}
\label{fig4}
  \end{figure}

It is hard to extract physical information from the coordinate-dependent
definition of the boundary curve. For example,
as long as the boundary curve is everywhere timelike, one can define null
coordinates $y^+ = x^+$ and $y^- = x^+_B(x^-)$ in which
the boundary curve is ``static", being given by the equation 
$y = 0$, where $y$ is the spatial coordinate $y \equiv (y^+ - y^-)/2$. 
On the other hand a quantity giving physical insight into the nature of 
the solutions is the outgoing 
radiation reaching future asymptotic null infinity, $\Im^+_R$. 
The outgoing radiation consists of a classical part (\ref{classb})  
that is reflected from the boundary, 
and a quantum part (\ref{tquant}). We calculate both in the 
manifestly asymptotically flat null coordinates on $\Im^+_R$.
In terms of the $u$ coordinate defined in Eq. (\ref{vu}) we have on $\Im^+_R$, 
\be \label{classTuu}
T_{uu}^{cl}(u) = (\lll x^- + P_+/\lll)^2 T_{--}^{cl}(x^-(u))
\ \ .
\ee
On the other hand, Eq. (\ref{tquant}) gives 
\be \label{quantTuu1}
\langle T_{uu} \rangle (u) = -\kappa  t_u (u)
\ \ , 
\ee
where we used the fact that $\rho$ and its derivatives vanish 
on $\Im^+_R$ in the manifestly asymptotically flat coordinates.
Furthermore, using the coordinate transformation (\ref{vu}) and Eq. 
(\ref{trelations}) with $x^+ \to x^-$ and $v \to u$, 
one can express $t_u (u)$ in terms of $t_-(x^-(u))$
to get
\be \label{quantTuu2}
\langle T_{uu} \rangle (u) = (\kappa \lll^2/4) [1 - \lll^{-2}( \lll x^- 
+ P_+/\lll)^2 t_-(x^-(u))]
\ \ .
\ee 
Putting the numerically integrated function $q(z)$ (defined in Eq. 
(\ref{dimlesscoord})) and the initial data, $T^{cl}_{++} (x^+)$, in 
(\ref{classb}) and (\ref{tcondition}), we find the profiles of the outgoing 
matter fluxes $T_{uu}^{cl}$ and $\langle 
T_{uu} \rangle$. The results are shown in Fig.\ \ref{fig4} (a) and (b) in which 
the infalling matter is described by the profile given in Eq.
(\ref{profile}) with $\lll \e = 5$. 
In Fig.\ \ref{fig4} (a), the mass is $M/\lll = 0.1$ (which is 10\% of $a$), while in 
Fig.\ \ref{fig4} (b) it is $M/\lll = 1$. 
The solid curves depict the total stress-tensor of the outgoing radiation, 
$T^f_{uu}=T^{cl}_{uu} + \langle T_{uu} \rangle$, in units of $\lll^2$, 
while the gray curves correspond to the quantum radiation stress-tensor, 
$\langle T_{uu} \rangle$.  
Since $\kappa = 10^{-6}$ is much smaller than $M/\lll$, 
the quantum radiation is scaled up in the figures by a 
factor of $2/\kappa = 2 \times 10^{6}$ and hence does not show up in 
the solid curve. 



As we will see later in section V, when the total mass and the 
local density of the infalling matter are large enough, 
then a black hole is formed; 
otherwise the infalling matter escapes to infinity without forming 
a black hole. For a fixed width $\epsilon$ in Eq. (\ref{profile}), 
the mass of the infalling matter $M$ fully determines its profile.
Let $M_{cr}$ be the mass above which 
a black hole is formed. As discussed in section V, for
$\lll \epsilon = 5$, we find that $M_{cr}/\lll \simeq 1.5$.
In Fig.\ \ref{fig4} (a), we take the mass of the infalling matter to be 
about 6\% of the critical mass $M_{cr}$. We see that the total outgoing 
radiation is very similar to the incoming radiation, Fig.\ \ref{fig3}a. 
The profile of the matter field is almost unaffected by the 
evolution. The curvature and coupling are everywhere small and the 
infalling matter is just reflected from the boundary with almost no 
distortion. In the 4D case, this corresponds to ``weak'' initial data 
\cite{Christodoulou} for which the null (scalar) field goes through the 
origin of the radial coordinate, $r=0$, without developing large densities, 
and escapes to infinity with very little distortion. 

The quantum part of the outgoing radiation stress-tensor
$\langle T_{uu} \rangle $ in Fig.\ \ref{fig4}a is almost symmetric. 
(This is true only for a
symmetric incoming classical profile far from the critical point.)
The quantum radiation includes a region of negative-energy density
that insures energy conservation. 
In section IV we will calculate the quantum correlation function and show
that the negative-energy radiation is strongly correlated with the 
positive-energy radiation. This strong correlation is necessary for the 
quantum state on $\Im^+_R$ to be pure. 

In Fig.\ \ref{fig4} (b) the mass of the infalling matter is about 65\% of the 
critical mass $M_{cr}$. We see that the outgoing classical radiation 
is more distorted compared to that in Fig.\ \ref{fig4} (a). In this case the 
infalling matter strongly distorts the spacetime 
before getting reflected from the boundary and escaping to infinity without 
forming a black hole. Likewise the quantum radiation in Fig.\ref{fig4}(b) is 
no longer symmetric. The reason for 
this is the following: As $M$ increases, the total amount of 
positive-energy quantum radiation increases.
Because of energy conservation also the total amount of negative-energy 
radiation increases with $M$. While the width of the positive-energy radiation
increases with $M$, the width of the negative-energy  
radiation, $\Delta u$,  satisfies the quantum inequality \cite{BPP2,Ford}
  \be \label{quantuminequality}
|E_{neg}| \Delta u < \kappa \; , 
  \ee  
where $E_{neg}$ is the total amount of negative-energy quantum radiation 
and $\Delta u$ is its width (as measured by an asymptotic observer). 
Since $|E_{neg}|$ increases with $M$, the width $\Delta u$ must decrease and 
we get a non-symmetric profile of $\langle T_{uu} \rangle$. As we will see in 
section V, as $M$ approaches $M_{cr}$ the width $\Delta u$ approaches zero, 
and we get a brief burst of negative energy.  

As a consequence of general covariance of the effective action,
Eq. (\ref{Seff}), we have $\nabla^{\mu} T^f_{\mu \nu} = 0$. We find numerically 
that the energy defined in terms of $T^f_{\mu \nu}$ is conserved, i.e., 
$\int_{\Im^-_R} T^f_{vv} dv = \int_{\Im^+_R} T^f_{uu} du$. This is what one 
would expect if the reflecting boundary at $X \equiv \exp(-2\phi) = 0^+$ 
corresponds to a static boundary like $r=0$ in 4D, as discussed in the 
paragraph after Eq. (\ref{tcondition1}). The $\hbar$ expansion implies that also 
$\nabla^{\mu} T^{cl}_{\mu \nu} = 0$ and $\nabla^{\mu} \langle T_{\mu \nu} \rangle
= 0$ (the latter was used in arriving at $\langle T_{\mu \nu} \rangle$ 
from the trace anomaly). 
Nevertheless, we find numerically that the energies associated with the 
classical and quantum parts alone are not separately conserved. For example, 
$\int_{\Im^-_R} T^{cl}_{vv} dv \neq \int_{\Im^+_R} T^{cl}_{uu} du$. 
Evidently, because the total energy-momentum tensor, $T^f_{\mu \nu}$, determines 
$\phi$, and hence the boundary, it behaves like a static boundary only for 
$T^f_{\mu \nu}$ but not for $T^{cl}_{\mu \nu}$ or $\langle T_{\mu \nu} \rangle$ 
individually. Far below criticality, the classical and quantum 
parts of the energy 
are nearly conserved separately. However, as one approaches the 
critical solution the classical and quantum parts of the energy 
each are strongly non-conserved.

Next we consider the profile described by 
  \be \label{homogen}
T^{cl}_{vv}(v) = \left\{ \begin{array}{cccc}
                          0 & , & & v < 0 \\
    {M \over (v_0 + \e/2)} \sin^2(\pi v / \epsilon)
                            & , & & 0 < v < \epsilon/2 \\
   {M \over v_0 + \e/2} & , 
& & {\e \over 2} < v < v_0 + {\e \over 2}  \\
   {M \over (v_0 + \e/2)} 
\sin^2[\pi (v - \lll v_0) / \epsilon]
                          & , & & v_0+{\epsilon\over 2} < v < 
                                      v_0+\epsilon \\
                          0 & , & & v > v_0+\epsilon \; . 
                         \end{array} \right. 
  \ee 
In Fig.\ \ref{fig3}(b) we show the profile (\ref{homogen}) with $v_0/\e = 100$,
which is nearly a square waveform. 
For $v_0 \gg \e$, as shown in Fig.\ \ref{fig3}(b), the energy-momentum tensor 
(\ref{homogen}) describes infalling matter of nearly homogeneous density. 
The total width of the profile is $(v_0 + \e)$.
In terms of the $q$-coordinate defined in Eq. (\ref{dimlesscoord}), 
the profile (\ref{homogen}) is 
  \be \label{qqhomogen}
T^{cl}_{qq}[q] = \left\{ \begin{array}{cccc}
                          0 & , & &\ln(q) < 0 \\
    {M \over \lll^2 (v_0 + \e/2)q^2} \sin^2[\pi \ln(q) / (\lll \epsilon)]
                            & , & & 0 < \ln(q) < \lll \epsilon/2 \\
    {M \over \lll^2(v_0 + \e/2) q^2} & , 
& & \lll \epsilon /2 < \ln(q) < \lll(v_0+{\epsilon\over 2}) \\
   {M \over \lll^2 (v_0 + \e/2)q^2} 
\sin^2\{\pi [\ln(q) - \lll v_0] / (\lll \epsilon)\}
                          & , & & \lll(v_0+{\epsilon\over 2}) < \ln(q) < 
                                      \lll(v_0+\epsilon) \\
                          0 & , & & \ln(q) > \lll(v_0+\epsilon) \; . 
                         \end{array} \right. 
  \ee 
We take (\ref{qqhomogen}) with $\lll v_0 = 10$ and $\lll \e = 0.1$, and 
integrate (\ref{ODE}) for different values of $M/\lll$. 
Using the numerically integrated function, $ q(z)$, we 
calculate the outgoing classical and quantum radiation on $\Im^+_R$,  
shown in Fig.\ \ref{fig4} (c) and (d). In (c) we take $M/\lll = 0.1$ and in (d) 
$M/\lll = 2$. Because the mass distribution of (\ref{homogen}) 
is more spread out than that of (\ref{profile}), $M_{cr}$ is larger,
having the value $M_{cr}/\lll \simeq 3.0$.
In these figures, the solid curves describe the total radiation 
while the gray curves describe the quantum radiation scaled up by $2 \times 
10^6$. In Fig.\ \ref{fig4}c the mass is about 3\% of the critical mass, which is the 
case of weak initial data, and is similar to Fig.\ \ref{fig4}a. The total outgoing 
radiation is very similar to the incoming classical radiation, 
Fig.\ \ref{fig3}(b), and the quantum radiation is spread across the outgoing 
classical radiation symmetrically. In Fig.\ \ref{fig4}d the mass is about 65\% 
of the critical mass, as was the case in Fig.\ \ref{fig4}b. The outgoing radiation 
is more distorted and the quantum radiation is non-symmetric. 

\begin{figure}
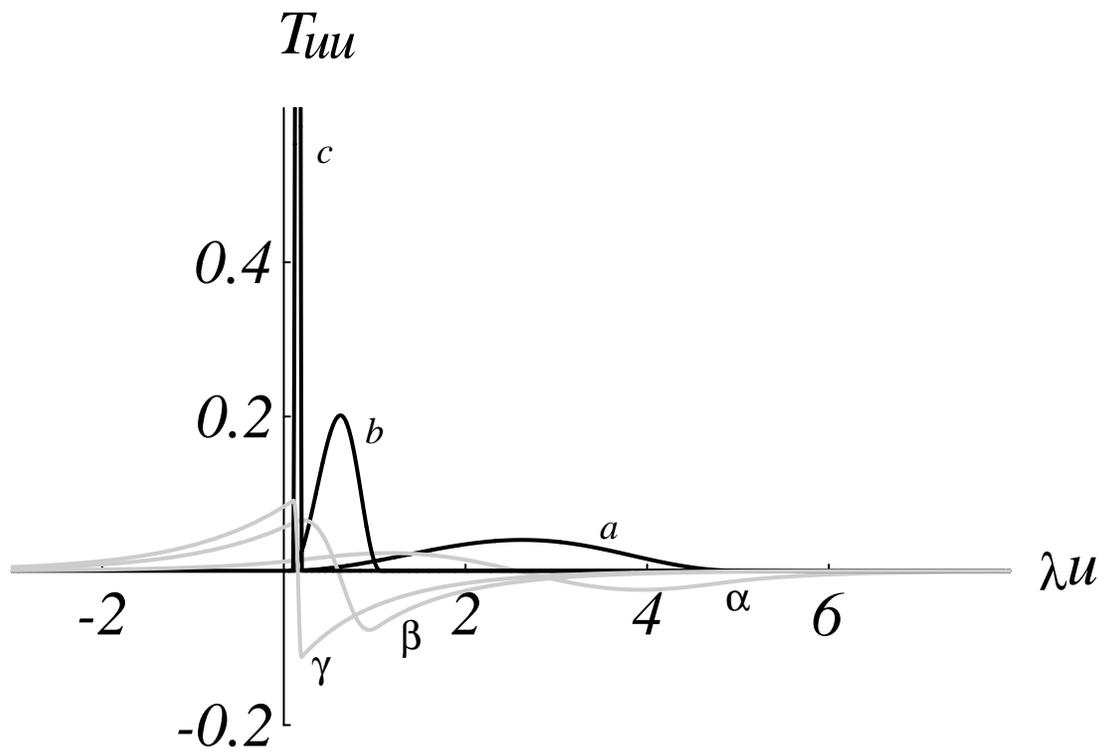
 
\caption{The stress tensor $T_{uu}$ describing the 
outgoing radiation for different values of $\lambda \epsilon$ 
in the case of the infalling matter shown in Fig. 3(a) with 
$M/\lambda = 0.1$. See the text for more details.} 
\label{fig5}
  \end{figure}

Finally, recall that in Ref.~\cite{BPP2} the case in which the classical 
matter had the form of an incoming shock-wave was studied.
To recover the results of Ref.~\cite{BPP2}, 
consider profiles (\ref{profile}) with $M/\lll = 0.1$ but with different 
values of the width $\lll \e$. 
In Fig.\ \ref{fig5}, the curves $a$ and $\alpha$ are, respectively, the total  
and quantum outgoing radiation in the case $\lll \e = 5$. 
The curves $b$ and $\beta$ are the total and 
quantum outgoing radiation for $\lll \e = 1$, and $c$ and $\gamma$ correspond 
to $\lll \e = 0.1$. One can see that as 
$\lll \e$ becomes smaller, the (incoming and outgoing) classical 
radiation becomes very localized, while the transition of the quantum radiation 
from positive to negative-energy values becomes more sudden. 
In the limit $\lll \e \rightarrow 0$, the classical energy-momentum 
tensor can be described by a delta function $T^{cl}_{++} = 
(M/\lll x^+_0) \delta (x^+ - x^+_0)$ and one recovers the results 
of Ref. \cite{BPP2}. One should remember however that as $\lll \epsilon 
\rightarrow 0$, the derivative $p'$ in Eq. (\ref{quantb}) diverges, and the 
large $N$ limit should be taken in such a way that the second term on the
right-hand-side of Eq. (\ref{quantb}) can be neglected \cite{BPP2}. 


\section{Information}
\subsection{Classical Structure}
In section III we considered infalling matter with very little 
structure. In order to encode non-trivial information in 
the classical infalling matter one should consider more complicated profiles. 
For example one can encode information in bits. In this case the infalling 
null matter can be a sequence of pulses each corresponding to one bit of 
information. Let us send a message in binary numbers. 
The number ``$1$" is described by a pulse 
of relative height $1$ and ``$0$" by a pulse of relative height $0.5$. 
As an example consider the following incoming 
energy-momentum tensor   
  \be \label{Hi}
T^{cl}_{vv}(v) = \sum_{i=1}^{16} l_i \Theta_i(v) ,
  \ee
where
  \be
\Theta_i(v) = \left\{ \begin{array}{cccc}
 {20 M \over 11} \sin^2[10 \pi (v - i) ]
                            & , & & i < v < i + 0.1 \\
                    0       & , & & \mbox{elsewhere} \; , 
                      \end{array} \right. 
  \ee
and 
  \be 
l_i \in \{ 0.5,1,0.5,0.5,1,0.5,0.5,0.5,0.5,1,1,0.5,1,0.5,0.5,1 \} . 
  \ee             
In Fig.\ \ref{fig3}c we show the profile (\ref{Hi}). It corresponds to the 
binary number $0100100001101001$, which is the word ``Hi'' in ASCI. 
In the $q$-coordinate the profile (\ref{Hi}) is 
  \be \label{qqHi}
T^{cl}_{qq}[q] = \sum_{i=1}^{16} l_i \Theta_i(q) ,
  \ee
where
  \be
\Theta_i(q) = \left\{ \begin{array}{cccc}
 {20 M \over 11 \lll q^2} \sin^2\{10 \pi [\ln(q) - i] \}
                            & , & & i < \ln(q) < i + 0.1 \\
                    0       & , & & \mbox{elsewhere} \; . 
                      \end{array} \right. 
  \ee  
Using (\ref{qqHi}), we integrate (\ref{ODE}) for different 
values of $M/\lll$ and find $q(z)$. Then we 
calculate the classical and quantum radiation on $\Im^+_R$, shown  
in Fig.\ \ref{fig6}. In (a) we take $M/\lll = 0.1$, and in (b) we take 
$M/\lll = 2$. The solid curves describe the total 
outgoing radiation, while the gray curves describe the quantum 
radiation, scaled by $2/\kappa = 2 \times 10^6$.   



For the profile (\ref{Hi}), we find numerically that the critical 
mass is $M_{cr}/\lll \simeq 2.5$.
In Fig.\ \ref{fig6}a the mass is about 4\% of the critical mass, and indeed 
we see that the outgoing radiation is very similar to the incoming 
one, Fig.\ \ref{fig3}c. This case is similar to the ones in Fig.\ \ref{fig4}a and 
\ref{fig4}c. One can easily recover the ``classical information'', 
(the word ``Hi''), from the outgoing radiation in Fig.\ \ref{fig6}a. Also the 
quantum radiation is quite symmetric. We can see from Fig.\ \ref{fig6}a that 
the classical and quantum radiation are correlated. 
Actually one can recover the ``classical information'' from the 
quantum radiation alone. Each classical pulse results in a sharp 
decrease (``jump'') of the quantum radiation. We see that 
for a pulse of a relative height 1 the jump is 
twice as large as the jump for a pulse of a relative height 0.5. Therefore 
by looking at the different jumps in the quantum radiation one can 
recover the ``classical information''. 

In Fig.\ \ref{fig6}b the mass is about 80\% of the critical mass, and 
the outgoing radiation is much more distorted than in Fig.\ \ref{fig6}a. As in the 
cases shown in Fig.\ \ref{fig4}b and \ref{fig4}d, it is difficult to recover the
``classical information'' of Fig.\ \ref{fig3}c from the profile shown in 
Fig.\ \ref{fig6}b. 

\subsection{Quantum correlation function}

As discussed earlier in section II, we choose the quantum state to be
the vacuum on $\Im^-_R$, so that
the quantum contribution to the energy-momentum tensor 
on $\Im^-_R$ is always zero, $\langle T_{vv} \rangle (v) =0$. 
By construction (with the reflecting boundary 
conditions and no black hole formation), the evolution of the quantum
state must be unitary. How can we see that the quantum radiation 
on $\Im^+_R$ is described by a pure state? The created quantum radiation 
reaching $\Im^+_R$ before the escaping classical matter is 
the beginning of the Hawking radiation which is almost thermal
and by itself does not correspond to a pure state. 
A pure state can be recovered from the radiation on $\Im^+_R$  
only if there are strong correlations between 
the early-time Hawking radiation and late-time negative-energy  
radiation reaching $\Im^+_R$ after the escaping outgoing classical matter. 
We demonstrate these correlations by calculating the correlation function 
  \be \label{corr}
C_{\mu\nu,\mu'\nu'}(x,x') \equiv \langle \hat{T}^f_{\mu\nu}(x) 
\hat{T}^f_{\mu'\nu'}(x') \rangle - 
\langle \hat{T}^f_{\mu\nu}(x)\rangle \langle \hat{T}^f_{\mu'\nu'}(x') \rangle .
  \ee
The correlations between different points $u$ and $u'$ on $\Im^+_R$ are given 
by $C_{uu,u'u'}(u,u')$, where $u$ is the asymptotically flat null coordinate 
on $\Im^+_R$. For the reflecting boundary conditions, we have a closed 
form expression\footnote{In Ref. \cite{BPP2}, the factor $N \hbar^2$ did not 
appear explicitly in Eq. (6) of Ref. \cite{BPP2} and the values of 
$C_{--,--}(u,u')$ quoted there are in units of $N \hbar^2$.} \cite{Carlitz}
  \be \label{Cuu}
C_{uu,u'u'}(u,u') = {N \hbar^2\over 8\pi} {\left( \partial_u v_B(u) \right)^2 
\left( \partial_{u'} v_B(u') \right)^2 \over {\left( v_B(u) - v_B(u') 
\right)}^4 } , 
  \ee
where $v_B(u)$ is the boundary curve in the $(u, v)$ coordinates defined 
in Eq. (\ref{vu}). Using the numerical solution for $q(z)$, we numerically 
find the functions $v_B(u) = \lll^{-1} \ln\{q[z(u)]\}$ and 
$\partial_u v_B(u)$, and from there can obtain $C_{uu,u'u'}(u,u')$.

The denominator in Eq. (\ref{Cuu}) is a rapidly varying function of $u$ and
$u'$. By defining the relative correlation  
  \be \label{relativecorr}
{C(+) \over C(-)} \equiv { C_{uu,u'u'}(u,u') \over C_{u''u'',u'u'}(u'',u')}
\quad ,\quad \mbox{where} \quad \quad v_B(u) - v_B(u') = v_B(u') - v_B(u'') 
  \ \ ,
\ee 
we get a measure for the correlations that varies more slowly as a function
of $u$ and $u'$ and also is not proportional to $\hbar$, which we take to 
be very small in the large $N$ limit.
This relative correlation is plotted as a function of $u$, with $u'$ held
fixed. The numerical results for the relative correlations (\ref{relativecorr}) 
and the outgoing quantum radiation shown in Fig.\ \ref{fig4}a,  
when we fix $\lll u' = 0$, are shown in Fig.\ \ref{fig7}a. 
Since for $u < 0$ the quantum radiation in Fig.\ \ref{fig4}a 
is exponentially small, the correlation function $C(-)$, which describes the 
correlations between the radiation at $u=0$ and the radiation at 
$u < 0$, is approximately the vacuum correlation function (there are almost 
no created particles in this region). On the other hand, the 
correlation function $C(+)$ describes the correlations between the 
radiation at $u=0$ and the radiation at $u > 0$. Since most of the radiation
is in the region $u > 0$, we can regard $C(-)$ as a reference function 
relative to which $C(+)$ is being measured. 
If $C(+)/C(-) < 1$, then the correlations described by $C(+)$ 
are weaker than the vacuum correlations, while if $C(+)/C(-) > 1$, then
the correlations are stronger.  



We see from Fig.\ \ref{fig7}a that when $u$ is in the region of positive-energy 
radiation (see also Fig.\ \ref{fig4}a), we have $C(+)/C(-) < 1$. This is expected 
since the positive-energy radiation is the precursor of the uncorrelated
thermal Hawking radiation. On the other hand, for $u$ in the region of 
negative-energy radiation we have $C(+)/C(-) > 1$. Namely, the correlations 
between a point in the region of positive-energy radiation, i.e., at $u=0$, 
and a point in the region of negative-energy radiation are {\em stronger} 
than the vacuum correlations. 
It is just when the energy of the quantum radiation becomes negative 
that $C(+)/C(-)$ becomes greater then one. Hence, the negative-energy 
radiation is not only necessary for energy conservation, but is also 
instrumental in recovering the correlations present in the final pure state.
Before we can discuss what happens to information when $M$ is very near the 
critical mass $M_{cr}$, we must study the properties of the near-critical
spacetimes. 

\section{Nearly critical spacetimes}

\subsection{Approaching the critical solutions}
The reflecting boundary conditions can only be imposed on those sections of
the boundary curve that are timelike. Along these sections, the 
boundary curve can be written in the form $x^+ = x^+_B(x^-)$, where 
$x^+_B(x^-)$ is a well-defined function, and one can impose the 
condition (\ref{reflect}). Since the boundary curve is dynamical,  
the initial data determines its nature through the evolution equations.
If the boundary is everywhere timelike, then the general solution we found 
earlier is regular, energy-conserving, and unitary. Such a solution is called
a subcritical 
(i.e., a non-black hole) solution. If on the other hand the boundary becomes 
spacelike in some regions of spacetime, then the spacetime develops a 
spacelike singularity ~\cite{BPP1}, which is initially hidden behind an 
apparent horizon. Such a solution is called a supercritical solution and 
describes an evaporating black hole ~\cite{BPP1}. 
One can therefore distinguish between two regions in the space of solutions 
or the space of initial data: black hole and non-black hole solutions. 
Furthermore, one can find continuously varying parameters 
$p_i$, specifying the initial data, such that all solutions with $p_i<p_i^*$ 
are subcritical, while all solutions with $p_i>p_i^*$ are supercritical. 
The solutions on the boundary separating the two regions (i.e., the ones 
with $p_i=p_i^*$) are called the critical solutions. 
As a result of the non-linearity of 
the system, the physical properties of the solutions need not be continuous
functions of the parameters at the critical values $p_i^*$.

We would like to study the transition between the subcritical and supercritical 
regions in our 
semiclassical theory. Using our numerical integration we can easily determine 
whether the solution is subcritical or supercritical: As long as the boundary 
curve is timelike we can write the boundary equation, $x^+=x^+_B(x^-)$, 
in the inverse form $x^- = x^-_B(x^+)$, where $x^-_B$ is the
inverse function of $x^+_B$. 
Since the boundary is smooth, the quantity $\partial x^-_B / \partial x^+$ is 
continuous. For a timelike boundary $\partial x^-_B /\partial x^+$ 
is positive, for a spacelike boundary $\partial x^-_B /\partial x^+$  
is negative, and for a null boundary $\partial x^-_B /
\partial x^+ = 0$. So at the point where the initially timelike boundary 
becomes null, just
before becoming spacelike, the derivative  $dq/dz$ in Eq. (\ref{ODE}) 
(which is the inverse of $\partial x^-_B/\partial x^+$) {\em diverges} 
and the numerical integration terminates. Therefore, all solutions for which 
the integration terminates at a finite value of $x^+_B$ are supercritical, 
while the others are subcritical. 

For the profiles in Fig.\ \ref{fig3} the continuous dimensionless parameters in  
solution space are: $M/\lll$ and $\lll\e$ in case (a), 
$M/\lll$, $\lll\e$ and $\lll v_0$ in case (b), 
and $M/\lll$ (the $l_i$ are not continuous parameters) in case (c). 
We approach the critical solutions by varying one of the parameters, while
keeping the others fixed. First we take $m \equiv M/\lll$ to be the 
free parameter. In case (a) we fix $\lll \e 
= 5$, and in case (b) we fix $\lll \e = 0.1$, $\lll v_0 = 10$. 
We find that in all the above cases there exists a critical value $m^*$
such that if $m<m^*$ the solutions are subcritical while if $m>m^*$ the 
solutions are supercritical. The value of $m^*$ depends on the values of 
the fixed parameters. To study the behavior of the solutions just below 
criticality, we numerically integrate the boundary equation for the cases 
$0 < \Delta m/m^* << 1$, where $\Delta m \equiv m^* - m$. 
We take $\kappa = 10^{-2}$ in this section, in order to be able 
to easily probe the regime, $\Delta m \leq \kappa$.  (The values of $m$
near $m^*$ are of order 1 and thus are still large with respect to
$\kappa$.) Then we calculate the 
outgoing radiation on $\Im^+_R$. In Fig.\ \ref{fig8} we show the  
radiation on $\Im^+_R$ for the three cases (a), (b) and (c)
of Fig.\ \ref{fig3}, where $\Delta m /m^* \sim 10^{-5}$. We shift the $u$-coordinate 
to $u - u_*$, where $u_*$ corresponds to the last classical reflected 
null ray, (i.e., $u_* = \lll^{-1}\ln[-\lll x^-_B(x^+_2) - P_+/\lll]$). 
In Fig.\ \ref{fig8}a we show the total stress tensor, $T^f_{uu}$, in units of 
$\lll^2$, describing the outgoing radiation on $\Im^+_R$, and in Fig.\ \ref{fig8}b 
we show only the quantum part. The quantum radiation is scaled 
by $4/(\kappa \lll^2)$, such that $\langle T_{uu} \rangle =1$ 
corresponds to the constant Hawking radiation from a 2D black hole 
\cite{CGHS}. 



We see from Fig.\ \ref{fig8} that while the early-time radiation strongly 
depends on the details of the classical infalling matter, the late-time 
radiation in the region $u>u_* - \lll^{-1}$ 
is almost the same for all the cases (a), (b) and (c). 
In the region $u > u_* - \lll^{-1}$ the classical part, $T^{cl}_{uu}$, increases
rapidly to large values of the order of  
$10^{8} \lll^2$, and then sharply 
decreases to zero. The quantum part, $\langle T_{uu} \rangle$, approaches 
the constant Hawking radiation, $\langle T_{uu} \rangle = \kappa \lll^2/4$,  
just before $u = u_*$. It then decreases rapidly 
to negative values of the order of $-10^{8} \lll^2$, and   
finally increases rapidly to zero. This late-time behavior appears to be 
independent of the profile of the infalling matter, as long as 
$0 < \Delta m / m^* << 1$. Similarly, if we approach a critical solution
by varying any one of the other parameters $p$, while holding the rest
fixed, we find the same late-time behavior as long as $0 < (p^* -p)/p^* \ll 1$.  
Let $T_{max}$ and $T_{min}$ 
be the maximum and minimum values of the total radiated 
energy flux $T^f_{uu}$ on $\Im^+_R$. For all the cases that we studied 
we find numerically that as $(p^* - p)/p^* \rightarrow 0^+$,  
the value of $T_{min}$ approaches $-T_{max}$ and $T_{max}$ tends to infinity.
In the next subsection we analyze this critical behavior.

\subsection{Breakdown of the semiclassical approximation ?}

To understand the above late-time behavior analytically,
we write $T^{cl}_{uu}(u)$ and $\langle T_{uu} (u) \rangle$ 
explicitly, using (\ref{classb}) and (\ref{quantTuu2}),  
  \be \label{classicalT}
T^{cl}_{uu}(u) = \left( {\partial x^- \over \partial u} 
\right)^2  \! \times \!
 \left( {\partial x^+_B(x^-) \over \partial x^-} \right)^2 
 \! \times \; T^{cl}_{++}(x^+_B) 
  \ee
and 
  \be \label{quantumT}
\langle T_{uu} (u) \rangle = {\kappa \lll^2 \over 4} \left[ 
1 - {1 \over (\lll x^+_B)^2} 
\left( {\partial x^- \over \partial u} \right)^2 
\left( {\partial x^+_B(x^-) \over \partial x^-} \right)^2 \right] . 
  \ee
We take the classical stress-tensor describing the incoming matter, 
$T^{cl}_{++}$, 
to be everywhere regular, i.e., finite and smooth. Therefore 
the non-trivial contributions to (\ref{classicalT}) 
and (\ref{quantumT}) come from the 
``redshift factor'', $(\partial x^- / \partial u)^2$, or from the ``blueshift 
factor'', $(\partial x^+_B / \partial x^-)^2$. 
Consider first the redshift factor. From the coordinate transformation 
(\ref{vu}) we get $(\partial x^-/\partial u)^2 = \exp(-2\lll u)$. 
If we ignore the back-reaction, i.e., take $\kappa = 0$ as
explained in \cite{BPP3}, the boundary curve of the critical solution 
first becomes null as $u \rightarrow 
\infty$, and the redshift factor approaches zero. However,  
when we include the back-reaction, the redshift factor at the point where 
the boundary becomes null is finite,
as can be seen as follows. 
Let $(x^+_c,x^-_c)$ be the point at which the boundary curve becomes null. 
At that point an apparent horizon is formed, since when the boundary 
curve becomes space-like (and describes a black hole singularity) it 
is surrounded by an apparent horizon ~\cite{BPP1}. 
The equation for the apparent horizon 
is $\partial e^{-2\phi} /\partial x^+ = 0$ ~\cite{RST1}, 
and using (\ref{Xsolution2}) we get 
  \be \label{apparenthorizon}
- \lll x^-_{\rm ah}(x^+) = P_+(x^+) / \lll + {\kappa \over 4\lll x^+} 
\ \ ,
  \ee
where the apparent horizon curve is $x^- = x^-_{\rm ah} (x^+)$.
Defining $\Delta P_+ = P_+ (x^+ \to \infty) - P_+(x^+_c)$ and 
using (\ref{vu}) and (\ref{apparenthorizon}), we find at $(x^+_c, x^-_c)$
  \be \label{redshiftfactor1}
{\left( {\partial x^- \over \partial u} \right)}^2_{x^-_c} = 
{\left( {\kappa \over 4\lll x^+_c} - {\Delta P_+\over \lll} \right)}^2 \; .
  \ee
Note that the apparent horizon is
null at $(x^+_c,x^-_c)$. This can be seen by showing that the derivative of 
(\ref{apparenthorizon}) with respect to $x^+$ vanishes at the point
$(x^+_c, x^-_c)$ on the boundary $X = 0$, where $X$ is given by 
Eq. (\ref{Xsolution2}). From the definition $P_+(x^+) = 
\int dx^+ T^{cl}_{++} (x^+)$ 
and the fact that $\partial x^-_{\rm ah} /\partial x^+ = 0$ at
$(x_c^+, x_c^-)$, we get 
  \be \label{Tppc}
T^{cl}_{++}(x^+_c) = {\kappa \over (2 x^+_c)^2} \; . 
  \ee
It follows that $T^{cl}_{v v}(v(x^+_c))$ is of order $\kappa \lll^2$,
which is the same as the quantum contribution. Therefore, $x^+_c$ must
be close to the cut-off point $x^+_2$ of the classical matter
distribution (the location of $x^+_2$ is shown in Fig.\ \ref{fig2}, assuming that
the profile $T^{cl}_{v v}(v)$ has no anomalously long tail. 
Expanding $\Delta P_+ \simeq (\partial P_+ / \partial x^+) \Delta x^+
= T^{cl}_{++} (x^+) \Delta x^+$, where $\Delta x^+ \equiv x^+_2 - x^+_c
\ll x^+_c$,
and using (\ref{apparenthorizon})-(\ref{Tppc}) we finally get 
  \be \label{redshiftfactor}
{\left( \partial x^- \over \partial u \right)}^2_{x^-_c} \simeq 
\left( {\kappa \over 4\lll x^+_c} \right)^2 \left( 1 - {2\Delta x^+
\over x^+_c} \right) \simeq \left( {\kappa \over 4\lll x^+_c} \right)^2 \; . 
  \ee
Thus the redshift factor (\ref{redshiftfactor}) is finite at $(x^+_c,x^-_c)$, 
and is zero only if we neglect the back-reaction ($\kappa = 0$). 

Next consider the blueshift factor, $(\partial x^+_B/\partial x^-)^2$. 
Obviously when the boundary curve becomes null the blueshift factor 
diverges. This divergence, together with the finiteness of the redshift 
factor and $T^{cl}_{++}$, is the reason for the divergences in 
(\ref{classicalT}) and (\ref{quantumT}) at $u_c \equiv u(x^-_c)$. 
But what about the total 
stress-tensor, $T^f_{uu} = T^{cl}_{uu} + \langle T_{uu} \rangle$? From 
(\ref{classicalT}), (\ref{quantumT}), and (\ref{Tppc}) we see that 
at $(x^+_c,x^-_c)$ the diverging terms in $T^{cl}_{uu}$ and $\langle T_{uu} 
\rangle$ cancel each other, leaving a finite $T^f_{uu}$: 
  \be \label{Ttotalc}
T^f_{uu}(u_c) = {\kappa \lll^2 \over 4} \; . 
  \ee
So why do we get divergences in the total stress tensor, shown in 
Fig.\ \ref{fig8}a? To understand this behavior, 
consider a point $(x^+_0,x^-_0=x^-_B(x^+_0))$ near $(x^+_c,x^-_c)$
on the boundary of a critical solution. Defining 
$\delta \equiv x^+_c - x^+_0$ and expanding (\ref{classicalT}), 
(\ref{quantumT}) about $(x^+_c,x^-_c)$ to leading orders in $\delta$, 
dropping the finite
term (\ref{Ttotalc}), we find 
  \be \label{divergingT}
T^f_{uu}(u_0) \simeq - \left( {\kappa\over 4 \lll x^+_c} \right)^2 \left[ 
{\kappa \over 2 (x^+_c)^3} + \left(\partial T_{++}^{cl}
\over \partial x^+\right)_{x^+_c} \right] {\delta \over (a \delta + 
b \delta^2)^2} \; , 
  \ee
where $a \equiv [\partial^2 x^-_B/\partial (x^+)^2]_c$, 
$b \equiv [\partial^3 x^-_B/\partial (x^+)^3]_c$, and $u_0=u(x^-_0)$. 
The point $(x^+_c,x^-_c)$ 
cannot be a local maximum of the boundary curve, since the boundary 
curve of the critical solution never becomes space-like. It also cannot 
be a local minimum since by construction it is the first (and only) point 
at which the boundary is null. Therefore the point $(x^+_c,x^-_c)$ 
must be an inflection point, i.e., $a = 0$. Also, for a general critical
solution we have $\left(\partial T_{++}^{cl} / \partial x^+ \right)_{x^+_c}
< - \kappa / [2 (x^+_c)^3]$, which is necessary for the apparent horizon
in (\ref{apparenthorizon}) to become timelike beyond the point $(x^+_c,x^-_c)$. 
We therefore get from (\ref{divergingT}) that $T^f_{uu}(u_0) \simeq A\delta^{-3}$ 
with $A > 0$. For points on the boundary to the past of $x^+_c$, 
$T^f_{uu} \simeq +A |\delta |^{-3}$ and for points to the future of $x^+_c$,
$T^f_{uu}\simeq -A|\delta |^{-3}$ as $|\delta| \rightarrow 0$. Thus for critical 
solutions the radiation flux diverges in opposite ways on both
sides as $x^+ \to x^+_c$. For almost critical 
solutions one gets the nearly divergent results shown in Fig.\ \ref{fig8}. 

The fact that $T^f_{uu}$ diverges does not necessarily signal a breakdown of the 
semiclassical approximation in which the metric and dilaton 
fields are treated as classical dynamical fields. 
From conservation of energy we know that the total amount of outgoing radiation 
(the sum of positive and negative parts), 
is finite and equals the total amount of incoming radiation. 
Moreover, the total amount of positive-energy radiation alone, 
$E_+$, in the region between $x^-_c$ and $x^-_0$ on 
$\Im^+_R$, is proportional to $T_{max} \Delta x^-$, 
where $\Delta x^- \equiv x^-_c - x^-_0$. From $x^+_c - x^+_0 = \delta$, we 
find that $\Delta x^- = b \delta^3$. It then follows from 
Eq. (\ref{divergingT}) that $E_+$ is finite. 
The same holds for the total amount of negative-energy radiation. 
Thus the divergence in the density, $T^f_{uu}$, can be 
viewed as describing a ``shock wave" or ``thunderpop" \cite{RST}. 
Although a shock wave does not necessarily imply 
a breakdown of the semiclassical approximation, an examination of the 
fluctuations in $\hat{T}^f_{\mu \nu}$ provides 
strong evidence for such a breakdown. Let us 
calculate the fluctuations in $ \hat{T}^f_{\mu \nu}$. 
As an estimate for the fluctuations one can use the quantity \cite{LFord,MikoRad}
  \be \label{fluctuations}
{\Delta'}_{\mu \nu \mu' \nu'}(x,x') \equiv \left| {\langle \hat{T}^f_{\mu \nu}(x) 
\hat{T}^f_{\mu' \nu'}(x') \rangle - 
\langle \hat{T}^f_{\mu \nu}(x) \rangle \langle \hat{T}^f_{\mu' \nu'}(x') \rangle 
\over \langle \hat{T}^f_{\mu \nu}(x) \rangle 
\langle \hat{T}^f_{\mu' \nu'}(x') \rangle } \right| = 
\left| {C_{\mu \nu , \mu' \nu'}(x,x') \over 
\langle \hat{T}^f_{\mu \nu}(x) \rangle 
\langle \hat{T}^f_{\mu' \nu'}(x') \rangle } \right| \; . 
  \ee
Far below criticality, with $x$ and $x'$ sufficiently apart, 
the numerator in Eq. (\ref{fluctuations}) will be less than or 
of the order of $N \hbar^2$ (see Eq. (\ref{Cuu})). 
On the other hand, the denominator 
is of the order of $\kappa^2 = N^2 \hbar^2$. Hence, in this case 
the fluctuations (\ref{fluctuations}) will be of the order of $1/N$ even in the 
region of negative energy. In the large $N$ limit these fluctuations are 
small and the semiclassical approximation is valid. 
However, for critical or nearly critical solutions this is no longer the case. 
Let us estimate ${\Delta'}_{uuu'u'}(u,u')$ for a critical or nearly critical 
solution, with $u$ and $u'$ sufficiently far apart that $v_B(u) - 
v_B(u') \sim \lll^{-1}$, which sets a natural length scale. 
Let $u'$ approaches $u_c$, and take $u$ to be in the region of negative 
energy such that $v_B(u) - v_B(u') \sim \lll^{-1}$, which is consistent with the quantum 
inequality (\ref{quantuminequality}). Then 
using (\ref{Cuu}), (\ref{quantumT}), (\ref{Ttotalc}) and (\ref{fluctuations}), 
we get for $u' \rightarrow u_c$  
  \be \label{Deltauc}
{\Delta'}_{uuu'u'}(u,u') \sim {1\over N}
{ \left( {\partial v_B(u') \over \partial u'} \right)}^2_{u' \rightarrow u_c} \; . 
  \ee
For any given $N$, no matter how large, while keeping $u$ fixed, 
we can find $u'$ sufficiently close to
$u_c$ such that the blueshift factor, $(\partial v_B / \partial u)^2$, 
in (\ref{Deltauc}) is dominant over $N$, and 
the fluctuations (\ref{Deltauc}) are large. 
We regard this anomalous behavior as implying a breakdown of the 
semiclassical approximation at criticality. 
Unlike in the 4D case, where the temperature of the black hole formed 
just above criticality approaches infinity, in the 2D case the black hole 
temperature is always a finite constant, $T_{BH} = \lll / 2\pi$, but 
nevertheless the semiclassical approximation (even in the large $N$ 
limit) breaks down at criticality. 
Also, for the critical solution the quantity 
$p'=\partial x^+_B/ \partial x^-$ diverges 
at $x^-_c$, and dropping the last term on the r.h.s. of 
Eq. (\ref{quantb}), i.e., the moving mirror term, becomes problematic 
even in the large $N$ limit. 

\section{High densities of information}
\subsection{Quantum limitations on the negative-energy radiation}
Let us calculate the relative-correlation function (\ref{relativecorr}) 
for nearly critical solutions. As in Fig.\ \ref{fig7}a, 
we take the infalling matter to be the one shown in 
Fig.\ \ref{fig3}a, with $\lll \e = 5$, but while in Fig.\ \ref{fig7}a we take 
$\Delta m / m \sim 0.9$, in this section we take $\Delta m / m \sim 10^{-5}$. 
The results for the relative correlations are shown in Fig.\ \ref{fig7}b. 
We see that in the region of the Hawking radiation ($0 < \lll u < 10.5$) 
the relative correlations are very small. Then in the region of the 
brief burst of negative energy, the relative correlations increase sharply 
to extremely high values of the order of $10^{11}$, and sharply decrease 
to one. It is this sharp increase in the relative correlations that 
insure a unitary evolution. If it were not for the negative-energy burst 
the correlations would be lost and the final state would be a mixed state. 

The information encoded in these correlations is very dense 
as the burst of negative energy is very localized (see Fig.\ \ref{fig8}a). 
To find an upper bound on the duration of the negative-energy burst,
$\Delta u$, we use the quantum inequality (\ref{quantuminequality}). 
To estimate the value of $|T^f_{uu}|$, we use the 
analytic expressions (\ref{classicalT}), 
(\ref{quantumT}) for a nearly critical solution. 
The quantity $|E_{\rm neg}|$ is approximately of the order of 
\be \label{Eneg}
|E_{\rm neg}| \sim |(T^f_{uu})_{\rm min}| \Delta u \simeq  (T^f_{uu})_{\rm max} 
\Delta u \; .
\ee
From (\ref{classicalT}) and (\ref{quantumT}) one can see that 
$(T^f_{uu})_{max}$ corresponds to the maximum 
of $\partial x^+_B / \partial x^-$. 
Let $(x^+_m,x^-_m)$ be the point on the boundary curve for 
which $T^f_{uu}$ is maximum. Using the boundary equation 
$X(x^+,x^-_B) = X_B$ and the solution (\ref{Xsolution2}) we find that at 
the maximum of $\partial x^+_B / \partial x^-$
  \be \label{Tatmax}
T^{cl}_{++}(x^+_m) = {\kappa \over 2 (x^+_m)^2} \; . 
  \ee
\noindent From (\ref{classicalT}), (\ref{quantumT}),
(\ref{redshiftfactor}), and (\ref{Tatmax}) we find that
for a nearly critical solution (for which 
$(\partial x^-_B / \partial x^+)_{\rm min}$ is very small) 
  \be \label{Tmax}
(T^f_{uu})_{max} \simeq {\kappa^3 \over 64 \lll^2 (x^+_m)^4} 
{\left( \partial x^-_B \over \partial x^+ \right)}^{-2}_{\rm min} \; . 
  \ee 
From (\ref{Eneg}), (\ref{Tmax}) and the quantum inequality (\ref{quantuminequality}), 
we get 
  \be \label{deltau}
\Delta u \leq {8 \lll(x^+_m)^2 \over \kappa} 
{\left( {\partial x^-_B \over \partial x^+} \right)}_{\rm min}  \; . 
  \ee
Since $x^+_m < x^+_2$ is finite, as one approaches the critical solution 
$(\partial x^-_B / \partial x^+)_{min}$ 
approaches zero and so does $\Delta u$. If the amount of information, 
$\Delta I$, carried by this burst of negative energy is finite, 
then as we approach the 
critical solution the information density, $\dot{I}=\Delta I/\Delta u$, 
diverges. In the next subsection we show that this is indeed the case.

\subsection{Entropy and Information}

The results regarding the quantum part of the energy-momentum tensor 
discussed in the previous sections may plausibly be interpreted as arising 
from the creation of particle-antiparticle pairs \cite{Parker}. 
The particles reach infinity, $\Im^+_R$, and give rise to the 
positive-energy (Hawking) radiation, while the antiparticles carrying 
negative energy are reflected from the boundary and give rise to the 
negative-energy radiation on $\Im^+_R$. If it were not for the 
negative-energy burst of radiation, the correlations between the particles 
and antiparticles, shown explicitly in Fig.\ \ref{fig7}, would be lost and 
the final state would be a mixed state with non-zero entropy. 
The entropy of this mixed state can be found using the fact that 
the spectrum of the outgoing positive-energy quantum radiation is almost 
thermal, as can be seen by calculating the corresponding Bogolubov 
coefficients \cite{Nelson}. The entropy of this thermal radiation 
is the dimensionless Boltzmann entropy \cite{Preskill} 
  \be \label{thermalentropy}
S_{\rm Bol} = 2 \hbar^{-1} \sqrt{\kappa} \int_{-\infty}^{u_0} 
\sqrt{{\cal E}(u)} du \; , 
  \ee
where ${\cal E}$ is the energy density of the thermal quantum radiation, 
${\cal E} = \langle T_{uu} \rangle$, 
and $u_0$ is the value of $u$ at which the thermal radiation ends. From 
Fig.\ \ref{fig8}b we see that as we approach the critical solution, 
the thermal quantum radiation is almost independent of the specific profile 
of the infalling matter. Therefore to get an estimate of 
(\ref{thermalentropy}) one can calculate it 
for the case of very localized infalling matter. In that case one 
finds analytically that ${\cal E}$ is given by \cite{BPP2} 
\be \label{calE}
{\cal E} = {\kappa \lll^2 \over 4}\left( 1 - 
                  {1\over (1 + \lll \Delta e^{\lll u})^2}\right)
\ \ ,
\ee
where $\Delta \equiv M/ (\lll^3 x^+_0)$, and $x^+ = x^+_0$ is the null
trajectory of the localized (shock-wave) infalling matter. Then the
Boltzmann entropy of the thermal radiation is
\be \label{Scr}
S_{\rm Bol} \simeq {N\over 6} \ln \left( {4 M \over \lll \kappa} \right) \; , 
  \ee
where we assume that $\ln(4M/ \lll \kappa) \gg 1$. 
The entropy (\ref{Scr}) is by definition the amount of information 
that is lost by ignoring correlations between the thermal radiation and
the later burst of negative energy,  $I_{\rm lost} = S_{\rm Bol}$. 
This information cannot be recovered from the thermal radiation before 
the arrival of the negative-energy burst. After the burst of 
negative energy all the correlations between the particles and 
antiparticles are restored, and we get a pure state of zero entropy. 
This is expected
since we impose reflecting boundary conditions and our spacetime has a 
trivial topology. Therefore, the information $\Delta I$ that is gained 
during the arrival of the burst of negative energy equals $I_{\rm lost}$, 
  \be \label{information}
\Delta I = I_{\rm lost} = S_{\rm Bol} \simeq 
{N\over 6} \ln \left( {4 M \over \lll \kappa} \right) \; . 
  \ee
As we approach the critical solution, $\Delta I$ remains finite.
Using (\ref{deltau}) and (\ref{information}) we get 
  \be \label{inforate}
\dot{I} = {\Delta I \over \Delta u} \geq {N \lll \over 48} 
\ln \left({4 M \over \lll \kappa} \right) {\kappa\over (\lll x^+_m)^2} 
{\left( \partial x^+_B \over \partial x^- \right)}_{\rm max}  \; . 
  \ee
As the critical solution is approached, $\dot{I} \rightarrow \infty$. 

Even though our space is one dimensional, the result (\ref{inforate}) 
suggests that the way information is transferred by the negative-energy 
burst is different from that by which information is transferred in 
{\em linear} channels. The theoretical upper-bound 
on the bulk of information flow in $N$ linear channels is \cite{Bekens}
  \be \label{linearchannels}
\dot{I}_{\rm linear} \leq {E \over 2 \pi \hbar} \ln(N) \; , 
  \ee
where the ``message'' is transferred with positive energy $E$. 
Suppose that this bound could be 
extended to negative energies, $E < 0$, by replacing $E$ in 
(\ref{linearchannels}) by $|E|$. From Eq. (\ref{calE}), we find
\be
|E| \simeq {\kappa \lll \over 4} \ln \left( {4M\over \kappa \lll}\right) \; .
\ee
Then the theoretical upper bound (\ref{linearchannels}) for our system 
would be
  \be \label{linearbound}
\dot{I}_{\rm linear} \leq {N \lll \over 96 \pi} \ln \left({4M \over \kappa \lll}
\right) \ln (N) \; . 
  \ee
For any given sufficientely large value of $N$ one can find a range of 
solutions just below criticality for which  
the semiclassical approximation remains valid (i.e., the fluctuations 
(\ref{Deltauc}) are small) while the bound (\ref{linearbound}) on 
$\dot{I}_{\rm linear}$ is exceeded by $\dot{I}$ in (\ref{inforate}). 
This is achieved by requiring that $\sqrt{N} >> 
(\partial v_B / \partial u)_{\rm max} > \ln(N) / 8\pi$. 

It may be better to view the information associated with the negative energy 
in terms of informationstorage rather than information transfer.
The brief burst of negative energy can 
be considered as a very localized configuration containing a finite amount 
of information that is traveling in space. As one approaches criticality 
the density of stored information becomes unbounded. This seems to be in 
agreement with other considerations regarding the unboundedness of information 
storage densities in quantum systems \cite{Landauer}.    

\section{Conclusions}

In this work we present and study in detail a theory of semiclassical 
2D dilaton gravity with reflecting (conformal) boundary conditions. 
This theory shares many of the features of spherically symmetric 
semiclassical gravity. In particular, massless scalar fields 
can collapse to form a black hole that evaporates. 
If the energy and energy density of the initial configuration of the 
scalar fields are below certain critical values, i.e., the subcritical 
or weak initial data case, then the scalar fields do not form a black hole, 
but instead are 
reflected from the boundary and escape to infinity. On the other hand, 
if the energy and energy density are above the critical values, i.e., the 
supercritical or strong initial data case, then the scalar fields 
collapse to form a black hole. This created black hole 
evaporates by emitting Hawking radiation. 
In a previous work \cite{BPP1} we found that such  
evaporation leads to an end-state geometry similar to that of 
a 4D semi-infinite throat. Here we study subcritical solutions, especially 
those near criticality, for general smooth initial data. 

By combining 
analytical and numerical techniques we investigate the detailed structure 
of the classical and quantum one-loop contributions to the 
outgoing radiation reaching asymptotic future null infinity, $\Im^+_R$. 
For the subcritical solutions, we find that before the reflected massless
classical matter fields reach $\Im^+_R$, positive-energy quantum radiation 
is observed at $\Im^+_R$ and continues as the classical matter reaches 
$\Im^+_R$. This positive-energy quantum radiation is
followed by a flux of negative-energy quantum radiation.

For profiles of incoming classical matter that encode information
in bits consisting of pulses of two different amplitudes, we calculate
how the encoded information appears in the outgoing classical profile 
on $\Im^+_R$. For the quantum 
radiation, the situation is more involved. The outgoing 
positive-energy quantum radiation by itself does not describe a pure state, 
but is strongly correlated with the negative-energy quantum radiation 
reaching $\Im^+_R$ at a later time. 
This negative-energy quantum radiation not only insures conservation of energy
but also restores the correlations necessary for a pure final quantum state.

The study of solutions just below and at criticality 
gives insight into the information puzzle in a unitary framework. 
As one approaches the critical solution 
the classical outgoing radiation becomes very distorted at late times. 
Part of that classical radiation reaches $\Im^+_R$ with a time-delay 
and has the form of an extremely dense brief pulse. 
As the critical solution is approached the late-time 
classical energy density becomes highly distorted and ultimately diverges, 
making it impossible to recover the complete classical information. 
Regarding the quantum radiation, as one approaches the critical 
solution the early-time radiation is indistinguishable from thermal 
Hawking radiation from a black hole, while 
the energy density of the late-time negative-energy radiation diverges, 
making recovery of the quantum correlations impossible. 
Although the late-time energy 
density of the outgoing radiation is infinite in the critical case, 
the total amount of energy is finite and conserved. 
Nevertheless, the semiclassical approximation breaks down because 
the fluctuations in $\hat{T}^f_{\mu \nu}$ become very large. 

The above analysis shows that the black hole phase transition (subcritical 
$\rightarrow$ supercritical) and the information puzzle are intimately 
related. At the critical solution there is an apparent 
breakdown of predictability. 
However, this breakdown of predictability is related to 
the breakdown of the semiclassical approximation. 
The same divergence in energy density that makes recovery of information and 
quantum correlations impossible evidently makes the semiclassical 
approximation invalid due to large fluctuations in $\hat{T}^f_{\mu \nu}$.  
We would like to stress that even though we take the large 
$N$ limit to enforce the validity of the semiclassical approximation 
for subcritical solutions, the semiclassical approximation nevertheless  
breaks down at criticality. This may suggest that the breakdown of the 
semiclassical approximation at the onset of black hole formation is a 
fundamental result independent of the specific model to be studied. This is 
supported by the fact that the crucial features discussed in Sec. V B, 
i.e., the finite redshift factor and infinite blueshift factor, seem 
to be independent of the explicit semiclassical model. 
Thus, based on the semiclassical approximation alone, it may not be possible 
to trace the fate of information and correlations at the onset of 
black hole formation. The full quantum 
theory appears crucial to understanding if the overall evolution can be 
unitary at and after the onset of black hole formation. 
Nevertheless, our results do show that as criticality is 
approached via subcritical solutions, information and correlations become 
essentially irrecoverable while unitary evolution is preserved. 

\bigbreak\bigskip\bigskip\centerline{\bf Acknowledgments}\nobreak
We thank B. Allen, L. Ford, J. Friedman, and T. Roman for helpful 
discussions, and R. Landauer for bringing to our attention 
the unboundedness of information storage densities in quantum systems. 
This work was supported 
by the National Science Foundation under Grant No. PHY 95-07740.

\newpage

  \begin{figure} 
\caption{The stress tensor $T_{uu}$ describing the 
outgoing radiation for the infalling matter shown in Fig. 3(c). 
The solid curves describe the total outgoing radiation, while the 
gray curves describe only the quantum part. 
In (a) the mass is about 4\% of the critical mass, while 
in (b) it is about 80\% of the critical mass.}
\label{fig6}
  \end{figure}
  \begin{figure} 
\caption{The relative correlation function of the outgoing quantum radiation 
on $\Im^+_R$. In (a) the mass is about 6\% of the critical mass, while in 
(b) it is 99.999\% of the critical mass.}
\label{fig7}
  \end{figure}
  \begin{figure}
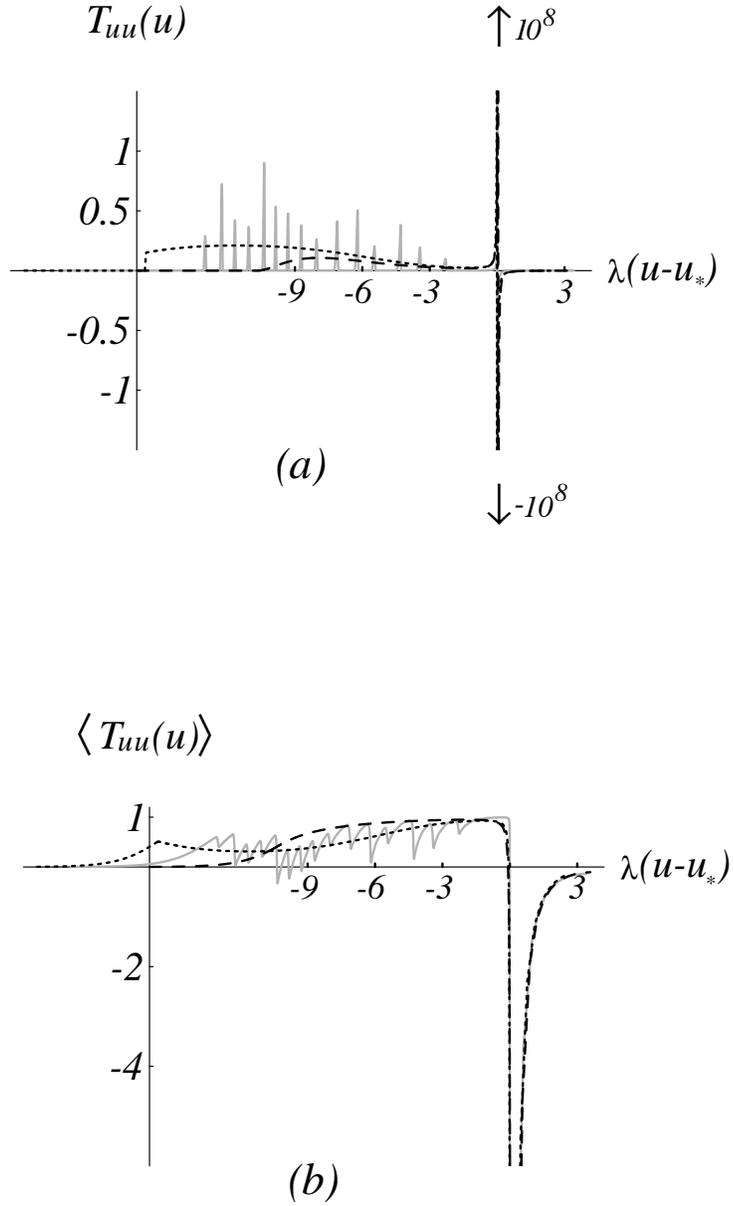
 
\caption{The stress tensor $T_{uu}$ describing the 
outgoing radiation for nearly critical solutions with $M$ 
being 99.999\% of the critical mass. In (a) we plot the total outgoing 
radiation and in (b) only the quantum part is plotted. 
The dashed, dotted and gray curves correspond to the 
infalling matter shown in Figs. 3(a), 3(b) and 3(c), respectively.}
\label{fig8}
  \end{figure}

\end{document}